\documentclass[12pt]{iopart}

\usepackage{iopams}
\usepackage{setstack}
\usepackage{graphicx}
\usepackage{slashbox}
\usepackage{mathptmx} 
\begin{document}

\title[Calculation of the anomalous magnetic moment of $\mu$]{Calculation of the anomalous magnetic moment of $\mu$ in
a model of electroweak-scale right-handed neutrinos}

\author{Jian-Ping Bu,Bei Jia}

\address{General Courses Department, Academy of Military Transportation, Tianjin 300161, China}
\ead{bjp1981@mail.nankai.edu.cn}
\begin{abstract}
The lepton flavor violating process $\mu\to e\gamma$ has been
previously studied in a model of electroweak-scale right-handed
neutrinos. We have calculated the decay amplitude in unitarity gauge \cite{Bu:2008fx} and $R_\xi$ gauge \cite{bujianping}. In this paper, we calculate the anomalous magnetic moment of $\mu$ in $R_\xi$ gauge.

\end{abstract}


\maketitle

\section{Introduction}
The Standard Model(SM) of elementary particle physics has been proved to be a remarkable successful description of high energy physics phenomena.However, the origin of neutrino mass is still an unknown fundamental problem\cite{Ramond:1999du,Kayser:2000ka}.In SM neutrinos are being introduced as a zero-mass.The discovery of flavor conversion of solar and atmospheric neutrinos\cite{Schwetz:2011qt,GonzalezGarcia:2010er} has established that neutrinos have nonzero mass and they mix among themselves, therefore providing the first evidence of new physics beyond the standard model.

On the other hand, there has been growing
interest in the anomalous magnetic moments of $\mu$ during the past few years.The most recent determination of $a_\mu$ in the SM is\cite{Roberts:2009qh}
\begin{eqnarray}
a_\mu^{\rm SM}=(11659183.4\pm4.9)\times10^{-10}.
\end{eqnarray}
The total SM prediction of $\mu$ differs from the experimental value\cite{Eidelman:1995ny}
\begin{eqnarray}
a_\mu^{\rm Exp}=(11659208.0\pm6)\times10^{-10}.
\end{eqnarray}
The difference between them is 3$\sigma$. Though 3$\sigma$ can not be regarded as the powerful proof of the new physics,it is possible that the difference will become larger with the development of experiments. So, we maybe get the proof of the new physics later.

A model with the above desired feature built in has been suggested
recently by Hung \cite{Hung:2006ap}. The model keeps the SM
gauge group albeit in a `vector-like' manner: Mirror fermions charged under SM gauge group, of
which right-handed neutrinos are a member, and two Higgs triplets
in a way that preserves the $\rho$ parameter to unity, plus a Higgs
singlet. In
particular, a right-handed neutrino which is sterile and are required to have a mass 
now becomes a member of a weak doublet of mirror leptons. A tiny
Dirac mass for neutrinos is offered by a scalar singlet whose
vacuum expectation value is not necessarily associated with the
electroweak scale, while a Majorana mass of order the electroweak
scale is provided by a scalar triplet. It is
conceivable that these new leptons could be discovered at high
energy colliders in the near future and the rich lepton flavor
structure could be observed in low energy processes. The weak charged
couplings are generally non-unitary with or without restricting to
the subspace of light leptons, and flavor changing neutral
currents (FCNC) occur in a way that is controlled by the weak
charged couplings.

The paper is organized as follows. First, we will introduce the Hung model briefly; for a full account of
the model, we refer to refs \cite{Hung:2006ap}. Then we show in
section 3 how to calculate the anomalous magnetic moment of $\mu$ in the virtual transition $\mu\to\mu\gamma$  in $R_\xi$ gauge, and demonstrate the final result. Our result is
summarized in the last section.
 
\section{Hung's Model}

We start with a brief description of the model relevant to our later analysis; for a full account of it, see Ref.\cite{Hung:2006ap}.
Considering three generations,the SM and mirror leptons with quantum numbers under the gauge group $SU(2)\times U(1)_Y$ are:
\begin{eqnarray}
F_L=\left(\begin{array}{c}n_L\\f_L\end{array}\right)~({\bf
2},Y=-1),&&f_R~({\bf 1},Y=-2);\nonumber\\
F^M_R=\left(\begin{array}{c}n_R^M\\f_R^M\end{array}\right)~({\bf
2},Y=-1),&&f_L^M~({\bf 1},Y=-2);
\end{eqnarray}
where the subscripts $L,~R$ refer to chirality and the superscript
$M$ to mirror, and the first number in parentheses stands for the
dimension of representation under the gauge group $SU(2)$. 
Because of anomaly cancelation,the quark sector also has mirror partners which are of no interest here.Besides the SM scalar doublet $\Phi$, the
model contains the new scalars
\begin{eqnarray}
\phi~({\bf 1},0),~\chi~({\bf 3},2),
\end{eqnarray}
plus an additional triplet $\xi~({\bf 3},0)$ that together with
$\chi$ preserves the custodial symmetry \cite{Chanowitz:1985ug} but
is irrelevant here.

The Yukawa couplings of leptons are, with the generation indices suppressed,
\begin{eqnarray}
-\calL_{\Phi}&=&y\overline{F_L}\Phi f_R+y_M
\overline{F^M_R}\Phi f_L^M+{\rm h.c.},\nonumber\\
-\calL_\phi&=&x_F\overline{F_L}F^M_R\phi+x_f\overline{f_R}f_L^M\phi
+{\rm h.c.},\nonumber\\
-\calL_{\chi}&=&\frac{1}{2}z_M \overline{(F^M_R)^C}(i\tau^2)\chi
F^M_R+{\rm h.c.}.
\end{eqnarray}
where $\psi^C=\calC\gamma^0\psi^*$,
$\calC=i\gamma^0\gamma^2$, and
\begin{eqnarray}
\chi=\frac{1}{\sqrt{2}}\vec{\tau}\cdot\vec{\chi}=\frac{1}{\sqrt{2}}
\left(\begin{array}{cc}\chi^+&\sqrt{2}\chi^{++}\\\sqrt{2}\chi^0&-\chi^+
\end{array}\right).
\end{eqnarray}
Note that a potential Majorana coupling of $\chi$ to $F_L$ is forbidden by imposing an appropriate $U(1)$ symmetry . Suppose the VEVs have the
structure:
\begin{eqnarray}
\langle\Phi\rangle=\frac{v_2}{\sqrt{2}}\left(\begin{array}{c}0\\1\end{array}\right),
~\langle\phi\rangle=v_1,
~\langle\chi\rangle=v_3\left(\begin{array}{cc}0&0\\1&0\end{array}\right),
\end{eqnarray}
where $v_{2,3}$ contribute to the masses of weak gauge bosons and
are naturally of order the electroweak scale while $v_1$ is not
necessarily related to it. In the basis of $f,~f^M$,the charged lepton mass terms are
\begin{eqnarray}
-\calL^f_{\rm m}&=&\left(\overline{f_L},\overline{f_L^M}\right)
m_f\left(\begin{array}{c}f_R\\f_R^M\end{array}\right)+{\rm
h.c.},\nonumber\\
m_f&=&\left(\begin{array}{cc}\displaystyle\frac{v_2}{\sqrt{2}}y& v_1x_F\\
v_1x^{\dagger}_f&\displaystyle\frac{v_2}{\sqrt{2}}y^{\dagger}_M%
\end{array}\right).
\end{eqnarray}
while the neutrino mass terms are
\begin{eqnarray}
-\calL^n_{\rm m}&=&
\frac{1}{2}\left(\overline{n_L},\overline{(n_R^M)^C}\right)m_n
\left(\begin{array}{c}n_L^C\\n_R^M\end{array}\right)+{\rm
h.c.},\nonumber\\
m_n&=&\left(\begin{array}{cc}0&v_1x_F\\v_1x_F^T&v_3z_M\end{array}\right).
\end{eqnarray}

We denote the mass eigenstate fields of the charged and neutral
leptons by $\ell_j$, $\nu_j$ respectively, with $j = 1, 2, 3$ for
those that are mostly ordinary leptons and $j = 4, 5, 6$ for those
that are mostly mirror. The seesaw mechanism operates for a Majorana mass of order the electroweak scale and Dirac proportional to $\nu_1$ which can be chosen small.This relaxes in some sense the tension in ordinary seesaw models between the generation of light neutrino mass and the observability of heavy neutrinos at colliders. The mass matrices are diagonalized by unitary transformations as follows
\begin{eqnarray}
\left(\begin{array}{c}
f_{L,R}\\
f_{L,R}^M\\
\end{array}\right)=X_{L,R}\ell_{L,R}, %
~\left( \begin{array}{c}
n_L^C  \\
n_R^M  \\
\end{array}\right)=Y\nu _R, %
~\left( \begin{array}{c}
n_L  \\
(n_R^M )^C  \\
\end{array} \right) = Y^ * \nu _{L},~\nu _{L}=\nu_R^C,\nonumber\\
\end{eqnarray}
where the unitary matrices $X_{L,R}$ and $Y$ satisfy
\begin{eqnarray}
X_L^\dagger  m_f X_R  = m_\ell = {\rm diag}(m_{\ell_i } ), %
~Y^T m_n Y = m_\nu   = {\rm diag}(m_{\nu _i } ),
\end{eqnarray}
with the mass eigenvalues $m_{{\ell_i},{\nu_i}}$ being real and nonnegative. There is a constraint on their masses from the zero texture,
\begin{eqnarray}
\sum\limits_{k =
1}^6{m_{\nu_i}}Y_{ik}Y_{jk}=0,\;\;\;\;\;\;{\rm for}\;i,j = 1,2,3.
\end{eqnarray}
Splitting each unitary matrix into two blocks that contain the
upper and lower rows respectively,
\begin{eqnarray}
X_{L,R}=\left(\begin{array}{c}X_{L,R}^{\rm u}\\
X_{L,R}^{\rm d}\end{array}\right),\;\;\;\;\;Y=\left(\begin{array}{c}Y^{\rm u}\\
Y^{\rm d}\end{array}\right),
\end{eqnarray}
the following relations will be required later:
\begin{eqnarray}
f_{L,R}=X_{L,R}^{\rm u}\ell_{L,R},~f_{L,R}^M=X_{L,R}^{\rm d}\ell_{L,R},~n_L=Y^{\rm u*}\nu_L,
~n_R^M=Y^{\rm d}\nu_R,
\end{eqnarray}
where the unitarity constraints become
\begin{eqnarray}
X_\alpha^{\rm u}X_\alpha^{\rm u*}=X_\alpha^{\rm d}X_\alpha^{\rm d*}=1_3,%
~X_\alpha^{\rm u}X_\alpha^{\rm d+}=0_3~({\rm for}~\alpha=L,R),\nonumber\\
Y^{\rm u}Y^{\rm u+}=Y^{\rm d}Y^{\rm d+}=1_3,~Y^{\rm u}Y^{\rm d+}=0_3,
\end{eqnarray}
and the zero texture constraint only acts on $Y^{\rm u}$:
\begin{eqnarray}
Y^{\rm u}m_\nu Y^{\rm u T}=0_3.
\end{eqnarray}

The above diagonalizing matrices will enter the gauge interactions of leptons.Some algebra yields,
\begin{eqnarray}
\calL_g&=&g_2\left(j^{+\mu}_WW_{\mu}^++j^{-\mu}_WW_{\mu}^-+J^\mu_ZZ_\mu\right)
+eJ^\mu_{\rm em}A_\mu,
\end{eqnarray}
where the currents are ($P_{L,R}=1\mp\gamma_5/2$)
\begin{eqnarray}
\sqrt{2}j^{+\mu}_W=\bar\nu\gamma^\mu\left(V_LP_L+V_RP_R\right)\ell,\nonumber\\
c_WJ^\mu_Z=\frac{1}{2}\overline{\nu}\gamma^\mu\left(
V_LV_L^\dagger P_L+V_RV_R^\dagger P_R\right)\nu%
-\frac{1}{2}\bar\ell\gamma^\mu\left(V_L^\dagger V_L P_L+
V_R^\dagger V_RP_R\right)\ell\nonumber\\
+s_W^2\bar\ell\gamma^\mu\ell,\nonumber\\
J^\mu_{\rm em}=-\bar\ell\gamma^\mu\ell,
\end{eqnarray}
with $\nu=\nu_R+\nu_L=\nu_R+\nu_R^C=\nu_L+\nu_L^C$,and $c_W=\cos\theta_W$,$s_W=\sin\theta_W$ with $\theta_W$ being the Weinberg angle.To relate the matrices $V_L$,$V_R$ to $X_a$ ($a=L,R$),$Y$,it is convenient to decompose the latter into the up and down $3\times6$ blocks,
\begin{eqnarray}
X_a=\left(\begin{array}{c}X_a^{\rm u}\\X_a^{\rm d}\end{array}\right),Y=\left(\begin{array}{c}Y^{\rm u}\\Y^{\rm d}\end{array}\right),
\end{eqnarray}
then
\begin{eqnarray}
V_L=Y^{{\rm u}T}X_L^{\rm u},V_R=Y^{{\rm d}\dagger}X_R^{\rm d},
\end{eqnarray}
with $V_L^TV_R=0$.These matrices are generally non-unitary and the deviation from unitarity induces FCNC in both sectors of neutrinos and charged leptons:
\begin{eqnarray}
&&V_LV_L^\dagger=Y^{{\rm u}T}Y^{{\rm u}*},~
V_RV_R^\dagger=Y^{{\rm d}\dagger}Y^{\rm d},\nonumber\\
&&V_L^\dagger V_L=X_L^{{\rm u}\dagger}X_L^{\rm u},~V_R^\dagger
V_R=X_R^{{\rm d}\dagger}X_R^{\rm d}.
\end{eqnarray}
The matrix blocks $X_L^{\rm d},~X_R^{\rm u}$ do not enter charged
current interactions since $f_L^M,~f_R$ are $SU(2)$ singlets. It is
important to note that the matrices in charged currents, $V_L,~V_R$,
are generally not unitary and there are flavor changing neutral
currents arising from the nonunitarity.

If we go on with our calculations, we will also need the Yukawa couplings of the
would-be Goldstone bosons (GB's). Although the original scalar
fields mix in a complicated manner via the terms in the scalar
potential, GB's are independent of mixing details due to gauge
symmetry and their Yukawa couplings only involve matrices that
already appear in gauge interactions. We list the Feynman rules as follows:
\begin{eqnarray}
\bar\nu_i\ell_\alpha G^+&:&+i\frac{g_2}{\sqrt{2}m_W}\left[
m_i(V^{L}_{i\alpha}P_L+V^{R}_{i\alpha}P_R)
-m_\alpha(V^{L}_{i\alpha}P_R+V^{R}_{i\alpha}P_L)\right],\nonumber\\
\bar\ell_\alpha \nu_iG^-&:&+i\frac{g_2}{\sqrt{2}m_W}\left[
m_i(V^{L*}_{i\alpha}P_R+V^{R*}_{i\alpha}P_L)
-m_\alpha(V^{L*}_{i\alpha}P_L+V^{R*}_{i\alpha}P_R)\right].
\end{eqnarray}
We give below other Feynman rules that will be
required later (all momenta incoming):
\begin{eqnarray}
A_\mu G^+(k_+)G^-(k_-)&:&i^2e(-i)(k_+-k_-)_\mu=ie(k_+-k_-)_\mu,
\nonumber\\
A_\mu W^\pm_\nu G^\mp&:&iem_Wg_{\mu\nu},\nonumber\\
A_\mu(k)W^+_\alpha(k_+)W^-_\beta(k_-)&:&\Gamma_{\alpha\beta\mu}
=g_{\alpha\beta}(k_--k_+)_\mu+g_{\beta\mu}(k-k_-)_\alpha\nonumber\\
&&+g_{\mu\alpha}(k_+-k)_\beta.
\end{eqnarray}

\section{Calculation in $R_\xi$ gauge} 
Now we begin to calculate the anomalous magnetic moment for $\mu$ in both W diagram and Z diagram.As discussed in \cite{bujianping}, there are two types of contributions at the
one loop level that are mediated respectively by charged current
interactions of $W^\pm$ and flavor changing neutral current
interactions of $Z^0$,corresponding to unitarity gauge and $R_\xi$ gauge.For the charged current diagrams we find that they involve a triple gauge coupling which is more divergent in the ultraviolet and the flavor mixing matrices in the charged current are not unitary.Thus in such a circumstance, it is highly desired that the calculation should be done in a safer $R_\xi$ gauge whose $\xi$ dependence is canceled as expected.

\begin{figure}
\includegraphics[width=16cm]{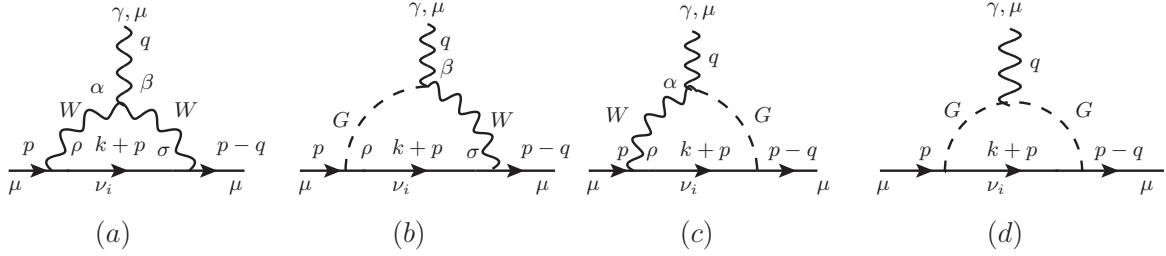} %
\caption{Diagrams that contribute to the decay $\mu\to\mu\gamma$.}
\label{fig1}
\end{figure}

\subsection{The contribution of W diagram}
First of all,we should calculate the amplitude for the decay $\mu ({p}) \to
\mu({p-q})\gamma(q,\epsilon) $ in W diagram whose Feynman diagrams are shown in Fig 1.The anomalous magnetic moment of $\mu$ is:
\begin{eqnarray}
a_\mu&=&a_\mu^W+a_\mu^Z\equiv\frac{g-2}{2}=F_2(q^2=0).
\end{eqnarray}
where
\begin{eqnarray}
\Gamma^\mu(p',p)\sim\frac{i\sigma_{\mu\nu}q^\nu}{2m_\mu}F_2(q^2).
\end{eqnarray}
The Lorentz and gauge symmetries dictate that the amplitude has
the following structure \cite{book}:
\begin{eqnarray}
\calA\sim\bar u_\mu i\sigma_{\lambda\nu}q^\nu
\left[x+y\gamma_5\right]\varepsilon^\lambda u_\mu,
\end{eqnarray}
so that we can concentrate on the $i\sigma_{\lambda\nu}q^\nu$ terms
to pick up the coefficients, $x$ and $y$.

We shall keep only terms up to the
linear order in the muon mass, $m_\mu$. These are indeed very good
approximations. The four  diagrams give the following on-shell  amplitudes;for a full presentation of our calculation details ,see Appendix:
\begin{eqnarray}
\fl T(a)=eg_2^2\bar u_\mu(i{\sigma _{\mu \nu }}{q^\nu })\frac{i}{{{{(4\pi
)}^2}}}\frac{1}{{m_W^2}}\Big\{ {m_\mu }A\Big[M(a) + P(a) \ln [r_i] +
Q(a)\ln [\xi ]\Big]\nonumber\\
+ {m_i}B\Big[M'(a) + P'(a)\ln [r_i] + Q'(a)\ln [\xi ]\Big]\nonumber\\
+m_\mu B\Big[M''(a) + P''(a)\ln [r_i]\Big]\nonumber\\
+m_\mu D\Big[M_D(a) + P_D(a)\ln [r_i] + Q_D(a)\ln [\xi ]\Big]\Big\}u_\mu,\\
\fl T(b)=eg_2^2\bar u_\mu(i{\sigma _{\mu \nu }}{q^\nu })\frac{i}{{{{(4\pi
)}^2}}}\frac{1}{{m_W^2}}\Big\{m_\mu A\big[M(b) + P(b) \ln
r_i+Q(b)\ln\xi\big]\nonumber\\
+{m_i}B\big[M'(b) + P'(b)\ln r_i + Q'(b)\ln\xi\big]\nonumber\\
+m_\mu D\big[M_D(b) + P_D(b)\ln r_i + Q_D(b)\ln\xi\big]\Big\}u_\mu,\\
\fl T(c)=eg_2^2\bar u_\mu(i{\sigma _{\mu \nu }}{q^\nu })\frac{i}{{{{(4\pi
)}^2}}}\frac{1}{{m_W^2}}\Big\{ {m_\mu }A\Big[M(c) + P(c) \ln r_i +
Q(c)\ln \xi \Big],\nonumber\\
+ {m_i}B\Big[M'(c) + P'(c)\ln r_i + Q'(c)\ln \xi
\Big]+{m_\mu}D\Big[M_D(c) + P_D(c)\ln r_i\nonumber\\
+ Q_D(c)\ln \xi
\Big]\Big\}u_\mu,\\
\fl T(d)=eg_2^2\ubar_\mu(i{\sigma _{\mu \nu }}{q^\nu })\frac{i}{{{{(4\pi
)}^2}}}\frac{1}{{m_W^2}}\Big\{ {m_\mu }A\Big[M(d) + P(d) \ln r_i
+Q(d)\ln \xi \Big]\nonumber\\
+{m_i}B\Big[M'(d) + P'(d)\ln r_i + Q'(d)\ln \xi \Big]\nonumber\\
+{m_\mu}D\Big[M_D(d) + P_D(d)\ln r_i+ Q_D(d)\ln \xi
\Big]\Big\}u_\mu.
\end{eqnarray}
The loop functions  such as $M(a),M(b),M(c),M(d)....$ and other functions are listed in Appendix.

Collecting the contributions from the four graphs yields the final answer:
\begin{eqnarray}
\fl a_\mu^W=\frac{\sqrt 2 G_F}{(4\pi)^2}\Big[m_\mu^2A_W\calF_W(r_i)
  +m_\mu m_iB_W\calG_W(r_i)+m_\mu^2 B_W\calP_W(r_i)+m_\mu^2
  D_W\calQ_W(r_i)\Big],\nonumber\\
\fl A_W=\sum_i(V^\ast_{Li\mu}V_{Li\mu}+V^\ast_{Ri\mu}V_{Ri\mu})=D_W,\nonumber\\
\fl B_W=\sum_i(V^\ast_{Li\mu}V_{Ri\mu}+V^\ast_{Ri\mu}V_{Li\mu}).
  \end{eqnarray}
where $\frac{G_F}{\sqrt 2}=\frac{g_2^2}{8m_W^2}$.The loop functions are:
\begin{eqnarray}
\calF_W&=&\frac{1}{{6{{(1 - r_i)}^4}}}(10 - 43r_i +
78{r_i^2} - 49{r_i^3} + 4{r_i^4} + 18{r_i^3}\ln r_i),\nonumber\\
\calG_W&=&\frac{1}{{{{(1 - r_i)}^3}}}( - 4 + 15r_i - 12{r_i^2} + {r_i^3} +
6{r_i^2}\ln r_i),\nonumber\\
\calP_W&=&\frac{1}{(1-r_i)^3}\left[13r_i^2-24r_i+11+2r_i(5-6r_i)\ln r_i\right],\nonumber\\
\calQ_W&=&\frac{1}{6(1-r_i)^4}\left[7-34r_i+33r_i^2-10r_i^3+4r_i^4-18r_i^2\ln r_i\right] .
\end{eqnarray}
It is good to see that the $\xi$ dependence is completely cancelled as
expected. 

\subsection{The contribution of Z diagram}
\begin{figure}
\includegraphics[width=16cm]{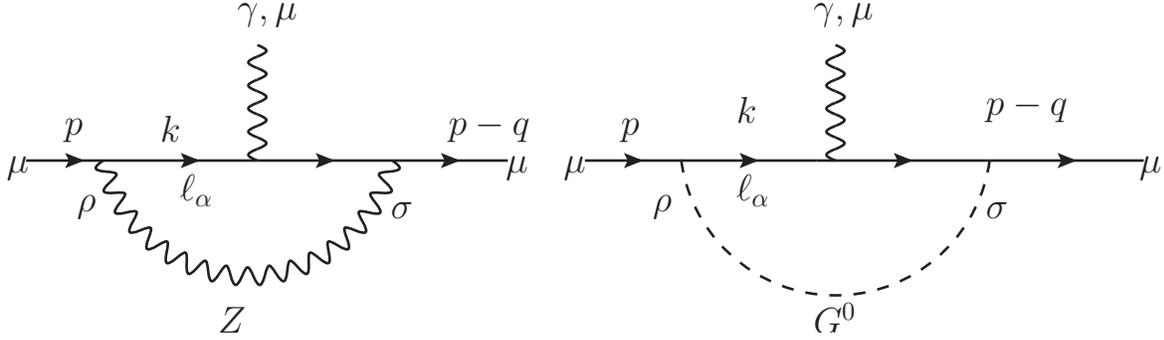} %
\caption{Diagrams that contribute to the decay $\mu\to\mu\gamma$.}
\label{fig2}
\end{figure}
We begin to calculate the amplitude of diagram Z whose Feynman diagrams are shown in Fig 2.We can use the same method in the diagram W.Thus we can give the contributions from the two graphs in Z diagram;for a full presentation of our calculation details ,see Appendix:
\begin{eqnarray}
\fl T\left[(b)_Z\right]=\frac{eg_2^2}{4c_Wm_Z^2}\frac{i}{(4\pi)^2}\ubar_\mu(i{\sigma _{\mu \nu }}{q^\nu })\Big\{m_\mu A'\big[MM(b) + PP(b) \ln
r_\alpha+QQ(b)\ln\xi\big]\nonumber\\
 +m_\alpha B'\big[MM'(b) + PP'(b)\ln r_\alpha + QQ'(b)\ln\xi\big]\nonumber\\
+m_\mu D'\big[MM_D(b) + PP_D(b)\ln r_\alpha + QQ_D(b)\ln\xi\big]\Big\}u_\mu,\\
\fl T[b_{G^0}]
=\frac{eg_2^2}{4c_Wm_Z^2}\frac{i}{(4\pi)^2}\ubar_\mu(i{\sigma _{\mu \nu }}{q^\nu })\Big\{m_\mu A'\big[MM(c) + PP(c) \ln
r_\alpha+QQ(c)\ln\xi\big]\nonumber\\
 +m_\alpha B'\big[MM'(c) + PP'(c)\ln r_\alpha + QQ'(c)\ln\xi\big]\nonumber\\
+m_\mu D'\big[MM_D(c) + PP_D(c)\ln r_\alpha + QQ_D(c)\ln\xi\big]\Big\}u_\mu,
\end{eqnarray}

Using (31) and (32),we finally get the contribution of the diagram Z to the anomalous magnetic moment of $\mu$:
\begin{eqnarray}
\fl a_\mu^Z=\frac{\sqrt{2}G_Fc_W}{(4\pi)^2}\Big[
  m_\mu^2A^\prime_Z\calF_Z(r_\alpha)+m_\mu m_\alpha B^\prime_Z\calG_Z(r_\alpha)+m_\mu^2
  B^\prime_Z\calP_Z(r_\alpha)+m_\mu^2 D^\prime_Z\calQ_Z(r_\alpha)\Big],\nonumber\\
\fl A^\prime_Z=\sum_\alpha(c^{L\ast}_{\alpha\mu}c^L_{\alpha\mu}
  +c^{R\ast}_{\alpha\mu}c^R_{\alpha\mu})=D^\prime_Z,\nonumber\\
\fl  B^\prime_Z=\sum_\alpha(c^{L\ast}_{\alpha\mu}c^R_{\alpha\mu}
  +c^{R\ast}_{\alpha\mu}c^L_{\alpha\mu}).
\end{eqnarray}
where
\begin{eqnarray}
\calF_Z&=&\frac{1}{{4{{(1 - r_\alpha)}^4}}}\left[-2+2r_\alpha-r_\alpha^2+10r_\alpha^3-9r_\alpha^4 + 2r_\alpha^2(-5+8r_\alpha)\ln r_\alpha\right],\nonumber\\
\calG_Z&=&\frac{1}{{{{2(1 - r_\alpha)}^3}}}\left[ -4-r_\alpha+8r_\alpha^2-3r_\alpha^3-6r_\alpha^2\ln r_\alpha\right],\nonumber\\
\calP_Z&=&\frac{1}{(1-r_\alpha)^3}\left[6r_\alpha^2-13r_\alpha+11+6r_\alpha(5-6r_\alpha)\ln r_\alpha\right],\nonumber\\
\calQ_Z&=&\frac{1}{12(1-r_\alpha)^4}\left[-14-6r_\alpha+57r_\alpha^2-54r_\alpha^3+18r_\alpha^4-6r_\alpha^2\ln r_\alpha\right] .
\end{eqnarray}
We find the result is also nothing to do with $\xi$ as expected.

\section{Numerical analysis}
We have calculated the W and Z diagrams in $R_\xi$ gauge.An interesting technical point is in order. We can also work in unitarity gauge and it should be simplest.But in the W diagram there is more ultraviolet divergent due to the triple gauge coupling.Thus there is no guarantee in this case that the order of removing the ultraviolet regulator commutes with that of taking the unitarity gauge limit.As a matter of fact, although the diagram is convergent in both unitarity and $R_\xi$ gauges, there is a finite difference in the terms linear in the lepton masses between the results obtained in the two gauges.
This caveat is restricted to the mentioned terms because terms of a higher order are convergent enough to allow the
free interchange of taking the limits.Considering this,we should work in $R_\xi$ gauge make sure not to have any problems.

Using (29) and (33),we finally obtain the anomalous magnetic moment of $\mu$:
\begin{eqnarray}
a_\mu&=&a_\mu^W+a_\mu^Z\equiv\frac{g-2}{2}=F_2(q^2=0).
\end{eqnarray}
The above anomalous magnetic moment involves several neutrino masses,the unknown $\alpha$ mass and many mixing matrix elements.In
our later numerical analysis, we shall make some approximations.First,the light neutrinos can
be safely treated as massless in the diagram W. Then, $\calF_W=\calF_W^l\to\frac{5}{3}$,$\calP_W=\calP_W^l\to11$,$\calQ_W=\calQ_W^l\to\frac{7}{6}$.The term $\calG$ can be dropped because of $m_i$.Second, for the heavy neutrino mass we choose $m_h=200,300$ up to 900 GeV.We use $m_W=80.2$ GeV.Thus $r_i=6.21887$.Then,$\calF_W=\calF_W^h\to0.0883211$,$\calG_W=\calG_W^h\to-0.350906$,$\calP_W=\calP_W^h\to-0.320226$,$\calQ_W=\calQ_W^h\to0.920466$. 

As a bonus of the
approximations, the anomalous magnetic moment of $\mu$ in the diagram W depends on the products of matrix
elements summed over light and heavy neutrinos respectively:
\begin{eqnarray}
&&V^l_1=\sum_{i=1}^3(V^\dagger_L)_{ei}(V_L)_{i\mu},~
V^l_2=\sum_{i=1}^3(V^\dagger_R)_{ei}(V_R)_{i\mu},\nonumber\\
&&V^l_3=\sum_{i=1}^3(V^\dagger_R)_{ei}(V_L)_{i\mu},~
V^l_4=\sum_{i=1}^3(V^\dagger_L)_{ei}(V_R)_{i\mu},
\end{eqnarray}
and similarly for $V^h_{1,2,3,4}$ with $i$ summed over $4,5,6$.

\begin{table}
\caption{\label{tabone}The anomalous magnetic moment of $\mu$($a_\mu\times10^{-13}$) is increasing with the increase of the heavy neutrino mass when the unknown $\alpha$ mass is a fixed value at first,while it is decreasing when the mass of the heavy neutrino is over 700 GeV.}

\small
\begin{tabular}{|l|*{9}{c|}}\hline
\backslashbox{$m_\alpha$(GeV)}{$m_i^h$(GeV)}&200 &300&400&500&600&700&800&900\\\hline
200&4.58086&4.58396 &4.58459&4.58476&4.58481&4.58482&4.58481&4.58480 \\\hline
300&5.68684&5.68994&5.69056 &5.69074&5.69079 &5.69080&5.69079&5.69078\\\hline
400&6.94412&6.94722&6.94784&6.94802&6.94807 &6.94807&6.94807&6.94805\\\hline
500&8.26680&8.26990&8.27052&8.27070&8.27075 &8.27075&8.27075&8.27073\\\hline
600&9.62381&9.62691&9.62753&9.62771&9.62776&9.62777&9.62776&9.62775\\\hline
700&11.0017&11.0048&11.0054&11.0056&11.0056 &11.0056&11.0056&11.0056\\\hline
800&12.3934 &12.3965&12.3971 &12.3973&12.3973 &12.3974&12.3973&12.3973\\\hline
900&13.7950&13.7981&13.7987&13.7989&13.7989 &13.7989&13.7989&13.7989\\\hline 
\end{tabular}

\end{table}

We find an algebraically simple case after some inspection.
Suppose the upper-right $3\times 3$ block of $Y$ is real. In this
scenario , our special neutrino spectrum (three almost massless
plus three almost degenerate and heavy) implies that the two
off-diagonal $3\times 3$ blocks of $Y$ vanish, the lower-right
block is trivially identity and the upper-left one is unitary.
Then, $V_1^l=(x_L^\dagger x_L)_{e\mu}$, $V^h_2 =-(x_R^\dagger
x_R)_{e\mu}$ while all others vanish, where $x_{L,R}$ are the
upper-left $3\times 3$ blocks of $X_{L,R}$ respectively. Since we
have no idea of their magnitudes, we sample randomly the real and
imaginary parts of $V_1^l,~V_2^h$ between $-2\times 10^{-6}$ and
$+2\times 10^{-6}$.

Thus for numerical analysis we can assume $A_W,D_W\sim10^{-6}$,$B_W\sim10^{-6}$ with $i$ summed over 1,2,3 while $B_W=0$ with $i$ summed over 4,5,6. We also use $\alpha=1/137.04$,$m_\mu=0.1056$ GeV.

We proceed to the diagram Z.First ,we can also assume the unknown $\alpha$ mass $m_\alpha=200,300$ up to 900 GeV.Then,$r_\alpha=4.81048$.We use $m_Z=91.187621$ GeV from PDG.Thus ,we get :$\calF_Z\to-3.15976$,$\calG_Z\to2.28058$,$\calP_Z\to6.91432$,$\calQ_Z\to1.90094$.We also make some approximations:
\begin{eqnarray}
\left(c^{L\ast}_{\alpha\mu}c^L_{\alpha\mu}\right)_\ll&\approx&2(V_L^\dagger V_L)_{\alpha\mu}^\ast(V_L^\dagger V_L)_{\alpha\mu}-4s_W^2(V_L^\dagger V_L)_{\alpha\mu}\sim2(1-2s_W^2)10^{-6}\nonumber\\
&\sim&10^{-7}\sim\left(c^{R\ast}_{\alpha\mu}c^R_{\alpha\mu}\right)_\ll\sim\left(c^{L\ast}_{\alpha\mu}c^R_{\alpha\mu}\right)_\ll,
\end{eqnarray}
Thus we obtain the different value of the anomalous magnetic moment of $\mu$ which has been shown in Table 1.

It is interesting that with the fixed unknown $m_\alpha$ the anomalous magnetic moment of $\mu$ is increasing with the increase of the heavy neutrino mass at first,while it is decreasing when the mass of the heavy neutrino is over 700 GeV.

\section{Conclusion} %
The subject of the anomalous magnetic moments for the muon is an exciting and fascinating topic because it represents the best compromise between sensitivity to new degrees of freedom describing physics beyond the standard model and experimental feasibility.
We have
calculated in $R_\xi$ gauge the anomalous magnetic moment of $\mu$ in a model suggested
recently. Because of the rich flavor structure of the model, the
weak charged currents involve nonunitary mixings between the neutral
and changed leptons and contain both left-handed and right-handed
chiralities.It is
conceivable that these new leptons could be discovered at high
energy colliders in the near future, while the rich lepton flavor
structure could be observed in low energy processes.

\section{Acknowledgements}

I would like to thank Prof.Yi Liao of NANKAI University for suggesting the topic and for helpful discussions.

\clearpage
\appendix
\section{The calculation of the contribution from W diagram}
We start with diagram $(b)$:
\begin{eqnarray}
\fl (b)=\frac{eg_2^2}{2}\ubar_\mu\int\frac{d^4k}{(2\pi)^4} %
\frac{N_\mu\left[ m_i\Big((V_{L})_{i\mu}P_L+(V_{R})_{i\mu}P_R\Big)-m_\mu\Big((V_{L})_{i\mu}P_R+(V_{R})_{i\mu}P_L\Big)\right]}{[k^2-\xi
m^2_W][(k+q)^2-m^2_W]}u_\mu,\nonumber\\
\end{eqnarray}
where
\begin{eqnarray}
\fl N_\mu=\gamma_\sigma\left[(V^\dagger_L)_{\mu i}P_L+(V^\dagger_R)_{\mu i}P_R\right]
\frac{1}{\kslash+\pslash-m_i}\left(g^{\sigma}_\mu-(1-\xi)\frac{(k+q)_\mu
(k+q)^\sigma}{(k+q)^2-\xi m^2_W}\right).\nonumber\\
\end{eqnarray}
To simplify $N_\mu$, we apply $\ubar_\mu\qslash=\ubar_\mu(-m_\mu+\pslash)$; i.e., we
can make the replacement, $\qslash\to\pslash-m_\mu$, when $\qslash$ is on
the far left. Note that the term $q_\mu$ has no contribution.Now $N_\mu$ can be split up as follows :
\begin{eqnarray}
N_\mu&=&{\rm I}+{\rm II}+{\rm III}+{\rm VI},
\end{eqnarray}
where
\begin{eqnarray}
{\rm I}&=&\left[(V^\dagger_L)_{\mu i}P_R+(V^\dagger_R)_{\mu i}P_L\right]\gamma_\mu\frac{\kslash+\pslash+m_i}{k^2+p^2-m_i^2},\nonumber\\
{\rm II}&=&-(1-\xi)\left[(V^\dagger_L)_{\mu i}P_R+(V^\dagger_R)_{\mu i}P_L\right]\frac{k_\mu}{(k+q)^2-\xi m^2_W},\nonumber\\
{\rm III}&=&-(1-\xi)(V_L^\dagger P_R+V_R^\dagger P_L)m_i\frac{k_\mu(\kslash+\pslash+m_i)}{\big[(k+p)^2-m_i^2\big]\left[(k+q)^2-\xi m_W^2\right]},\nonumber\\
{\rm IV}&=&(1-\xi)(V_L^\dagger P_L+V_R^\dagger P_R)m_\mu\frac{k_\mu(\kslash+\pslash+m_i)}{\big[(k+p)^2-m_i^2\big]\left[(k+q)^2-\xi m_W^2\right]}.
\end{eqnarray}
These three terms will be calculated separately below.

The term I gives the following contribution to the loop integral:
\begin{eqnarray}
\fl (b)_{\rm I}=\frac{eg_2^2}{2}\ubar_\mu\int\frac{d^4k}{(2\pi)^4} %
\frac{\left[(V^\dagger_L)_{\mu i}P_R+(V^\dagger_R)_{\mu i}P_L\right]\gamma_\mu(\kslash+\pslash+m_i)}{[k^2+p^2-m_i^2][k^2-\xi
m^2_W][(k+q)^2-m^2_W]}\nonumber\\
\times\left[ m_i\Big((V_{L})_{i\mu}P_L+(V_{R})_{i\mu}P_R\Big)-m_\mu\Big((V_{L})_{i\mu}P_R+(V_{R})_{i\mu}P_L\Big)\right]u_\mu,
\end{eqnarray}
As mentioned before, we only need pick up the
$(i{\sigma_{\mu\nu}}{q^\nu})$ terms. Using $\pslash u_\mu=m_\mu
u_\mu$, we can make the replacement $\pslash\to m_\mu$ when $\pslash$ is on the far
right. Since we work up to the linear order in $m_\mu$, it is
sufficient to expand to $O(p^2)$. The expansion yields several types
of terms. By symmetric integration, we get
$\gamma_\mu\kslash(-2k\cdot p)\to-\frac{1}{2}\gamma_\mu\pslash k^2$
which has no contribution to the desired Lorentz structure.
Similarly, the other two terms are, $\gamma_\mu\kslash(-2k\cdot
q)\to-\frac{1}{2}\gamma_\mu\qslash
k^2\to\frac{1}{2}(i\sigma_{\mu\nu}q^\nu)k^2$. The contribution of the I term is summarized as follows:
\begin{eqnarray}
\fl (b)_{\rm I}=\frac{1}{4}eg_2^2\bar u_\mu(i{\sigma _{\mu \nu }}{q^\nu })
m_iB\int_k
\frac{{{k^2}}}{{[{k^2}-m_i^2][{k^2}-\xi m_W^2]{{[{k^2} -
m_W^2]}^2}}}u_\mu
\nonumber\\
-\frac{1}{4}eg_2^2\bar u_\mu(i{\sigma _{\mu \nu }}{q^\nu })m_\mu
A\int_k
\frac{{{k^2}}}{{[{k^2}-m_i^2][{k^2}-\xi m_W^2]{{[{k^2}-
m_W^2]}^2}}}u_\mu,
\end{eqnarray}
where
\begin{eqnarray}
\int_k&\equiv&\int\frac{d^4k}{(2\pi)^4},\nonumber\\
A&=&\sum_i\left((V^\dagger_L)_{\mu i}(V_L)_{i\mu}P_R+(V^\dagger_R)_{\mu i}
(V_R)_{i\mu}P_L\right),\nonumber\\
B&=&\sum_i\left((V^\dagger_L)_{\mu i}(V_R)_{i\mu}P_R+(V^\dagger_R)_{\mu i}(V_L)_{i\mu}P_L\right).
\end{eqnarray}
We find that the term II has no contribution to $(b)$.

In calculating the term III,we will use: 
\begin{eqnarray}
&&k_\mu\kslash(k\cdot p)(k\cdot q)\nonumber\\
&\to&\frac{1}{48}(i{\sigma _{\mu \nu }}{q^\nu })\pslash k^4-\frac{1}{48}(i{\sigma _{\mu \nu }}{q^\nu })m_\mu^* k^4,\nonumber\\
&&k_\mu\kslash(k\cdot p)(k\cdot p)\nonumber\\
&\to&\frac{1}{24}(i{\sigma _{\mu \nu }}{q^\nu })\pslash k^4.
\end{eqnarray}
Note that adding the subscript * to $m_\mu$ means $P_L\to P_R,P_R\to P_L$ when $P_L$ and $P_R$ are on the far left.	Thus we get the contribution of the term III to the loop
 integral:
\begin{eqnarray}
\fl(b)_{\rm III}=\frac{1}{{24}}eg_2^2 \ubar_\mu  (i\sigma _{\mu \nu } q^\nu  )
m_i^2 m_\mu  A  (1 - \xi )\nonumber\\
\times\int_k k^4
\Big[- \frac{1}{{(k^2  - m_i^2 )^2 (k^2  - \xi m_W^2 )^3 (k^2  - m_W^2 )}}\nonumber\\
-\frac{1}{{(k^2  - m_i^2 )^2 (k^2  - \xi m_W^2 )^2 (k^2  - m_W^2 )^2 }}\nonumber\\
-2\frac{1}{{(k^2  - m_i^2 )^3 (k^2  - \xi m_W^2 )^2 (k^2  - m_W^2 )}}\Big]u_\mu\nonumber\\
+\frac{1}{8}eg_2^2 \bar u_\mu (i\sigma _{\mu \nu } q^\nu  )m_i^3 B  (1 - \xi )\int_k \frac{{k^2 }}{{(k^2  -
m_i^2 )^2 (k^2  - \xi m_W^2 )^2 (k^2  - m_W^2 )}}u_\mu\nonumber\\
+eg_2^2\ubar_\mu(i{\sigma _{\mu \nu }}{q^\nu })(1-\xi)m_i^2m_\mu D\int_k\Big[\frac{1}{24}\frac{k^4}{(k^2-m_i^2)^2(k^2-\xi m_W^2)^3(k^2-m_W^2)}\nonumber\\
+\frac{1}{24}\frac{k^4}{(k^2-m_i^2)^2(k^2-\xi m_W^2)^2(k^2-m_W^2)^2}\Big]u_\mu.
\end{eqnarray}
We can use the same method in the term IV and we get the contribution of it:
\begin{eqnarray}
\fl(b)_{\rm IV}=eg_2^2\ubar_\mu(i{\sigma _{\mu \nu }}{q^\nu })(1-\xi)m_i^2m_\mu D(-\frac{1}{8})\frac{k^2}{(k^2-m_i^2)^2(k^2-\xi m_W^2)^2(k^2-m_W^2)}u_\mu.
\end{eqnarray}
where
\begin{eqnarray}
D&=&\sum_i\left((V^\dagger_L)_{\mu i}(V_L)_{i\mu}P_L+(V^\dagger_R)_{\mu i}
(V_R)_{i\mu}P_R\right) .
\end{eqnarray}
Now we can get the contribution of diagram b to the loop integral:
\begin{eqnarray}
T(b)=eg_2^2\bar u_\mu(i\sigma_{\mu\nu}q^\nu)(m_\mu AF+m_i BG+m_\mu DW)u_\mu,
\end{eqnarray}
where
\begin{eqnarray}
\fl F=-\frac{1}{4}\frac{\xi }{{\xi  - 1}}I_1 +
\frac{1}{{24}}\frac{1}{{\xi  - 1}}\left(6 - \frac{r_i}{{r - 1}}\right)I_2 -
\frac{1}{{24}}\frac{r_i}{{\xi  - 1}}\left(\frac{1}{{\xi  - 1}} + 2\right)I_3
\nonumber\\
+\frac{1}{{24}}\frac{r_i}{{\xi-1}}\left(\frac{1}{{\xi  - 1}} -
\frac{1}{{r_i - 1}} + 2\right)I_4
+\frac{1}{{12}}\frac{r_i}{{(r_i - 1)(\xi  - 1)}}I_7
+ \frac{1}{{12}}r_i m_W^2 \xi I_9 \nonumber\\
+\frac{1}{{24}}r_i m_W^4 \xi ^2 I_{10}
- \frac{1}{{12}}r_i m_W^2 \left(1 - \xi
+ \frac{1}{{\xi  - 1}}\right)I_{11}
+ \frac{1}{{12}}r_i m_W^4 \xi ^2 I_{12},\nonumber\\
\fl G=\frac{1}{4}\frac{\xi }{{\xi  - 1}}{I_1} -
\frac{1}{4}\frac{1}{{\xi  - 1}}{I_2} + \frac{1}{8}\frac{r_i}{{\xi  -
1}}{I_3} - \frac{1}{8}\frac{r_i}{{\xi  - 1}}{I_4} -
\frac{1}{8}r_i{m_w}^2\xi {I_9},\nonumber\\
\fl W=\frac{1}{24}\frac{r_i}{(\xi-1)(r_i-1)}I_2+\frac{r_i}{\xi-1}\left[-\frac{1}{24}+\frac{1}{24(\xi-1)}\right]I_3\nonumber\\
+\frac{r_i}{\xi-1}\left[\frac{1}{24}-\frac{1}{24(\xi-1)}-\frac{1}{24(r_i-1)}\right]I_4+\frac{1}{24}m_W^2r_i\xi I_9-\frac{1}{24}m_W^4r_i\xi^2 I_{10}.\nonumber\\
\end{eqnarray}
Note that $r_i=m_i^2/m_W^2$,$I_n$ are integral functions listed in Appendix C.Substituting the $I_n$ functions
into $T(b)$ gives:
\begin{eqnarray}
\fl T(b)=\frac{i}{{{{(4\pi )}^2}}}\frac{1}{{m_W^2}}eg_2^2\bar
u_\mu(i{\sigma _{\mu \nu }}{q^\nu })\Big\{m_\mu A\big[M(b) + P(b) \ln
r_i+Q(b)\ln\xi\big]\nonumber\\
+{m_i}B\big[M'(b) + P'(b)\ln r_i + Q'(b)\ln\xi\big]\nonumber\\
+m_\mu D\big[M_D(b) + P_D(b)\ln r_i + Q_D(b)\ln\xi\big]\Big\}u_\mu,
\end{eqnarray}
where $M(b),P(b),Q(b),\dots$ are listed in Appendix E.

Now we proceed to diagram $(c)$ and $(d)$.Following the steps which are entirely similar to those in the
calculation of diagram $(b)$, we can obtain the contributions to the
amplitude from them:
\begin{eqnarray}
\fl T(c)=eg_2^2\bar u_\mu(i{\sigma _{\mu \nu }}{q^\nu })\frac{i}{{{{(4\pi
)}^2}}}\frac{1}{{m_W^2}}\Big\{ {m_\mu }A\Big[M(c) + P(c) \ln r_i +
Q(c)\ln \xi \Big],\nonumber\\
+ {m_i}B\Big[M'(c) + P'(c)\ln r_i + Q'(c)\ln \xi
\Big]+{m_\mu}D\Big[M_D(c) + P_D(c)\ln r_i\nonumber\\
+ Q_D(c)\ln \xi
\Big]\Big\}u_\mu,\\
\fl T(d)=eg_2^2\ubar_\mu(i{\sigma _{\mu \nu }}{q^\nu })\frac{i}{{{{(4\pi
)}^2}}}\frac{1}{{m_w^2}}\Big\{ {m_\mu }A\Big[M(d) + P(d) \ln r_i
+Q(d)\ln \xi \Big]\nonumber\\
+{m_i}B\Big[M'(d) + P'(d)\ln r_i + Q'(d)\ln \xi \Big]\nonumber\\
+{m_\mu}D\Big[M_D(d) + P_D(d)\ln r_i+ Q_D(d)\ln \xi
\Big]\Big\}u_\mu,
\end{eqnarray}
where $M(c),P(c),M(d),P(d),\dots$ are listed in Appendix F and G.

We finally come to the diagram (a) which is the most complicated one due to the appearance of a triple gauge coupling and double gauge boson propagators:
\begin{eqnarray}
\fl -\frac{2}{eg_2^2}(a)=\ubar_\mu\int\frac{d^4k}{(2\pi)^4}%
\gamma_\sigma\left[(V^\dagger_L)_{\mu i}P_L+(V^\dagger_R)_{\mu i}P_R\right]
\frac{1}{\kslash+\pslash-m_i}\gamma_\rho
\left[(V_L)_{i\mu}P_L+(V_R)_{i\mu}P_R\right]\Gamma_{\alpha\beta\mu}\nonumber\\
\times\frac{1}{k^2-m^2_W}\left[g^{\alpha\rho}- \frac{k^\alpha
k^\rho}{k^2-\xi m^2_W}(1-\xi)\right]
\frac{1}{(k+q)^2-m^2_W}\nonumber\\
\times\left[g^{\beta\sigma}- \frac{(k+q)^\beta
(k+q)^\sigma}{(k+q)^2-\xi m^2_W}(1-\xi)\right]u_\mu.
\end{eqnarray}
The above can be split into three terms corresponding to the product of the two propagators:
\begin{eqnarray}
-\frac{2}{eg_2^2}(a)=(gg)+(k)+(k+q)
\end{eqnarray}
where
\begin{eqnarray}
\fl(gg)=\ubar_\mu\int\frac{d^4k}{(2\pi)^4}%
\left[(V^\dagger_L)_{\mu i}P_R+(V^\dagger_R)_{\mu i}P_L\right]
\gamma^\beta\frac{1}{\kslash+\pslash-m_i}\gamma^\alpha
\left[(V_L)_{i\mu}P_L+(V_R)_{i\mu}P_R\right]\nonumber\\
\times\frac{\Gamma_{\alpha\beta\mu}}{[k^2-m^2_W][(k+q)^2-m^2_W]}u_\mu,\\
\fl-\frac{(k)}{1-\xi}=\bar u_\mu\int\frac{d^4k}{(2\pi)^4}%
\gamma^\beta\left[(V^\dagger_L)_{ei}P_L+(V^\dagger_R)_{\mu i}P_R\right]
\frac{1}{\kslash+\pslash-m_i}\kslash
\left[(V_L)_{i\mu}P_L+(V_R)_{i\mu}P_R\right]\nonumber\\
\times\frac{k^\alpha\Gamma_{\alpha\beta\mu}}{[k^2-m^2_W][k^2-\xi
m^2_W][(k+q)^2-m^2_W]}u_\mu,\\
\fl-\frac{(k+q)}{1-\xi}=\bar u_\mu\int\frac{d^4k}{(2\pi)^4}%
(\kslash+\qslash)\left[(V^\dagger_L)_{\mu i}P_L+(V^\dagger_R)_{\mu i}P_R\right]
\frac{1}{\kslash+\pslash-m_i}\gamma^\alpha
\left[(V_L)_{i\mu}P_L+(V_R)_{i\mu}P_R\right]\nonumber\\
\times\frac{(k+q)^\beta\Gamma_{\alpha\beta\mu}}
{[k^2-m^2_W][(k+q)^2-m^2_W][(k+q)^2-\xi m^2_W]}u_\mu.
\end{eqnarray}
The fourth term from the product of propagators has been discarded
since it does not contribute to the on-shell amplitude.

We start from the apparently easiest (actually the most complicated)
term $(gg)$. Simplify by aiming at the leading terms linear in
$m_\mu$:
\begin{eqnarray}
\gamma^\beta\frac{1}{\kslash+\pslash-m_i}\gamma^\alpha\Gamma_{\alpha\beta\mu}.
\end{eqnarray}
Note that the term $q_\mu$ and the term $\gamma_\mu$ can be dropped because they have no contribution to $(gg)$ and we use $\gamma_\alpha(\kslash+\pslash)\gamma^\alpha=-2(\kslash+\pslash)$.Thus we get:
\begin{eqnarray}
\fl\gamma^\beta\frac{1}{\kslash+\pslash-m_i}\gamma^\alpha\Gamma_{\alpha\beta\mu}\nonumber\\
\fl\to\frac{1}{(k+p)^2-m^2_i}\Big[ %
4\kslash k_\mu +2k_\mu(2\qslash-3m_i)
+\gamma_\mu\kslash(2\qslash-\pslash)
+\gamma_\mu 2\qslash(-\pslash+m_i) %
+m_i(-\qslash)\gamma_\mu\nonumber\\
-3m_\mu\qslash\gamma_\mu+4m_\mu k_\mu-m_\mu\kslash\gamma_\mu\Big].
\end{eqnarray}
Substitute (A.23) into (A.19) and again we should pick out
$(i\sigma_{\mu\nu}q^\nu)$ terms. By using the same method as in
calculating $(b)_{\rm I}$ we get the following contribution to the loop
integral:
\begin{eqnarray}
T(gg)&=&\bar u_\mu (i\sigma _{\mu \nu } q^\nu  )\Big[m_\mu AM+m_i
BN+m_\mu BH+m_\mu DS\Big]u_\mu,
\end{eqnarray}
where
\begin{eqnarray}
M&=&\frac{{4 - 13r_i + 9r_i^2 }}{{6(r_i - 1)^2 }}I_2
+ \frac{{( - 5r_i + 2)r_i}}{{3(r_i - 1)^2 }}I_4
+ \frac{2}{3}\frac{{r_i^2 }}{{(r_i - 1)^2 }}I_7
+ \frac{{3r_i - 1}}{{6(r_i - 1)^2 }}I_8,\nonumber\\
N&=&\frac{{3( - 2r_i + 1)}}{{2(r_i - 1)}}I_2  + \frac{{3r_i}}{{2(r_i
- 1)}}I_4,\nonumber\\
H&=&-3I_2+\frac{1}{2}\frac{r_i}{r_i-1}I_4-\frac{1}{2}\frac{1}{r_i-1}I_8,\nonumber\\
S&=&\frac{2r_i}{3(r_i-1)^2}I_2-\frac{r_i^2}{3(r_i-1)^2}I_4-\frac{1}{3(r_i-1)^2}I_8.
\end{eqnarray}
For (A.20), note that:
\begin{eqnarray}
k_\alpha\Gamma_{\alpha\beta\mu}&\to&-k_\mu(k+q)_\beta+g_{\beta\mu}(2q+k)\cdot k,
\end{eqnarray}
we simplify the structure:
\begin{eqnarray}
&&\gamma^\beta\left(V_L^\dagger P_L+V_R^\dagger P_R\right)\frac{k^\alpha\Gamma_{\alpha\beta\mu}}{\kslash+\pslash-m_i}\kslash\nonumber\\
&\to&{\rm I}''+{\rm II}''+{\rm III}''.
\end{eqnarray}
where
\begin{eqnarray}
{\rm I}''&=&-\left(V_L^\dagger P_R+V_R^\dagger P_L\right)m_i\frac{k_\mu(\kslash+\pslash+m_i)(m_i-\pslash)}{(k+p)^2-m_i^2},\nonumber\\
{\rm II}''&=&\left(V_L^\dagger P_L+V_R^\dagger P_R\right)m_\mu\frac{k_\mu(\kslash+\pslash+m_i)(m_i-\pslash)}{(k+p)^2-m_i^2},\nonumber\\
{\rm III}''&=&\left(V_L^\dagger P_R+V_R^\dagger P_L\right)\gamma_\mu(2k\cdot q+k^2)\frac{(\kslash+\pslash+m_i)(m_i-\pslash)}{(k+p)^2-m_i^2}.
\end{eqnarray}
Using the same method in diagram b we get the contribution of the above three terms:
\begin{eqnarray}
\fl T_{{\rm I}'}=(i{\sigma _{\mu \nu }}{q^\nu })\int_k\Big\{m_i^2m_\mu A\Big[-\frac{1}{12}\frac{k^4}{(k^2-m_i^2)^2(k^2-\xi m_W^2)(k^2-m_W^2)^3}\nonumber\\
-\frac{1}{6}\frac{k^4}{(k^2-m_i^2)^3(k^2-\xi m_W^2)(k^2-m_W^2)^2}\Big]-\frac{1}{4}m_i^3B\frac{k^2}{(k^2-m_i^2)^2(k^2-\xi m_W^2)(k^2-m_W^2)^2}\nonumber\\
+\frac{1}{12}m_i^2m_\mu D\frac{k^4}{(k^2-m_i^2)^2(k^2-\xi m_W^2)(k^2-m_W^2)^3}\Big\}.\nonumber\\
\fl T_{{\rm II}'}=(i{\sigma _{\mu \nu }}{q^\nu })m_i^2m_\mu D(-\frac{1}{4})\int_k\frac{k^2}{(k^2-m_i^2)^2(k^2-\xi m_W^2)(k^2-m_W^2)^2}.\nonumber\\
\fl T_{{\rm III}'}=(i{\sigma _{\mu \nu }}{q^\nu })\Big\{m_\mu A\Big[-\frac{1}{2}\frac{k^4}{(k^2-m_i^2)(k^2-\xi m_W^2)(k^2-m_W^2)^3}\nonumber\\
+\frac{1}{2}\frac{k^2}{(k^2-m_i^2)(k^2-\xi m_W^2)(k^2-m_W^2)^2}\Big]+m_i B\Big[\frac{1}{2}\frac{k^4}{(k^2-m_i^2)(k^2-\xi m_W^2)(k^2-m_W^2)^3}\nonumber\\
-\frac{1}{2}\frac{k^2}{(k^2-m_i^2)(k^2-\xi m_W^2)(k^2-m_W^2)^2}\Big]\Big\}.
\end{eqnarray}
Thus we can get:
\begin{eqnarray}
T\Big[-\frac{(k)}{1-\xi}\Big]=\bar u_\mu (i\sigma
_{\mu \nu } q^\nu  )\Big[m_\mu AM'+m_i BN'+m_\mu DW'\Big]u_\mu ,
\end{eqnarray}
where
\begin{eqnarray}
\fl M'=\frac{6  +  r_i -  12\xi+  4 r_i\xi}{{(r_i - 1)(1 - \xi )^2 }}I_2
 + \frac{{r_i\xi ^2 }}{{12(1 - \xi )^3 }}I_3
 + \frac{{r_i(2 + r_i - 3\xi  - 3r_i\xi  + 4r_i\xi ^2  - r_i^2 \xi ^2 )}}
 {{12(r_i - 1)^2 (1 - \xi )^3 }}I_4\nonumber\\
- \frac{{r_i(r_i + \xi  - 2r_i\xi )}}{{6(r_i - 1)^2 (1 - \xi )^2 }}I_7
+ \frac{{5r_i - 6}}{{12(r_i - 1)^2 (1 - \xi )}}I_8- \frac{{r_i\xi ^2 m_W^2 }}{{6(1 - \xi )^2 }}I_{11}, \nonumber\\
\fl N'=\frac{ - 2  -  r_i  +  4\xi-  r_i\xi}{{4(r_i - 1)(1 - \xi )^2 }}I_2
+ \frac{{r_i(\xi  - 2)}}{{4(1 - \xi )^2 }}I_3
+ \frac{{r_i(1 - r_i\xi )}}{{4(r_i - 1)(1 - \xi )^2 }}I_4 -
\frac{1}{{2(r_i - 1)(1 - \xi )}}I_8,\nonumber\\
\fl W'=\frac{-r_i(4-2r_i+r_i\xi-3\xi)}{12(r_i-1)^2(\xi-1)^2}I_2+\frac{r_i\xi(3-2\xi)}{12(\xi-1)^3}I_3+\Big[\frac{r_i(r_i\xi-1)}{4(r_i-1)(\xi-1)^2}\nonumber\\
-\frac{r_i(r_i+\xi-3r_i\xi+r_i^2\xi^2)}{12(r_i-1)^2(\xi-1)^3}\Big]I_4-\frac{r_i}{12(r_i-1)^2(\xi-1)}I_8.
\end{eqnarray}
Now we cope with the last term (A.21). Simplify first the combination:
\begin{eqnarray}
(\kslash+\qslash)\left(V_L^\dagger P_L+V_R^\dagger P_R\right)\frac{(k+q)^\beta\Gamma_{\alpha\beta\mu}}{\kslash+\pslash-m_i}
\gamma^\alpha
&\to{\rm I}(m_i)+{\rm II}(m_\mu).
\end{eqnarray}
where
\begin{eqnarray}
\fl{\rm I}(m_i)=\left(V_L^\dagger P_R+V_R^\dagger P_L\right)m_i\frac{-m_ik_\mu\kslash+k_\mu\kslash\pslash-m_i^2k_\mu+k^2\kslash\gamma_\mu+k^2\pslash\gamma_\mu+m_ik^2\gamma_\mu}{(k+p)^2-m_i^2},\nonumber\\
\fl{\rm II}(m_\mu)=-\left(V_L^\dagger P_L+V_R^\dagger P_R\right)m_\mu\frac{1}{\kslash+\pslash-m_i}(-k_\mu\kslash+\gamma_\mu k^2).
\end{eqnarray}
We can deal with the above two terms using the same method in diagram b.Thus we get the contribution of them to the loop integral:
\begin{eqnarray}
\fl T[{\rm I}(m_i)]
=(i{\sigma _{\mu \nu }}{q^\nu })\int_k\Big\{m_i^2m_\mu A\Big[-\frac{1}{12}\frac{k^4}{(k^2-m_i^2)^2(k^2-\xi m_W^2)^2(k^2-m_W^2)^2}\nonumber\\
-\frac{1}{12}\frac{k^4}{(k^2-m_i^2)^2(k^2-\xi m_W^2)(k^2-m_W^2)^3}-\frac{1}{6}\frac{k^4}{(k^2-m_i^2)^3(k^2-\xi m_W^2)(k^2-m_W^2)^2}\Big]\nonumber\\
+m_i^3B\frac{1}{4}\frac{k^2}{(k^2-m_i^2)^2(k^2-\xi m_W^2)(k^2-m_W^2)^2}\nonumber\\
+m_iB\Big[-\frac{1}{2}\frac{k^4}{(k^2-m_i^2)^2(k^2-\xi m_W^2)(k^2-m_W^2)^2}\nonumber\\
-\frac{1}{2}\frac{k^4}{(k^2-m_i^2)(k^2-\xi m_W^2)^2(k^2-m_W^2)^2}-\frac{1}{2}\frac{k^4}{(k^2-m_i^2)(k^2-\xi m_W^2)(k^2-m_W^2)^3}\nonumber\\
+\frac{k^2}{(k^2-m_i^2)(k^2-\xi m_W^2)(k^2-m_W^2)^2}\Big]\nonumber\\
+m_i^2m_\mu D\Big[\frac{1}{12}\frac{k^4}{(k^2-m_i^2)^2(k^2-\xi m_W^2)^2(k^2-m_W^2)^2}\nonumber\\
+\frac{1}{12}\frac{k^4}{(k^2-m_i^2)^2(k^2-\xi m_W^2)(k^2-m_W^2)^3}\Big]\Big\}.
\end{eqnarray}
\begin{eqnarray}
\fl T[{\rm II}(m_\mu)]
\to(i{\sigma _{\mu \nu }}{q^\nu })m_\mu D\Big[-\frac{1}{4}m_i^2\frac{k^2}{(k^2-m_i^2)^2(k^2-\xi m_W^2)(k^2-m_W^2)^2}\nonumber\\
+\frac{1}{2}\frac{k^4}{(k^2-m_i^2)^2(k^2-\xi m_W^2)(k^2-m_W^2)^2}+\frac{1}{2}\frac{k^4}{(k^2-m_i^2)(k^2-\xi m_W^2)^2(k^2-m_W^2)^2}\nonumber\\
+\frac{1}{2}\frac{k^4}{(k^2-m_i^2)(k^2-\xi m_W^2)(k^2-m_W^2)^3}-\frac{k^2}{(k^2-m_i^2)(k^2-\xi m_W^2)(k^2-m_W^2)^2}\Big].\nonumber\\
\end{eqnarray}
At last we obtain the following contribution to the loop integral:
\begin{eqnarray}
T\Big[-\frac{(k+q)}{1-\xi}\Big]= \bar u_\mu (i\sigma _{\mu \nu } q^\nu
)\Big[m_\mu AM''+m_i BN''+m_\mu DW''\Big]u_\mu,
\end{eqnarray}
where
\begin{eqnarray}
\fl M''=\frac{{r_i }}{{6(r_i  - 1)(1 - \xi )}}I_2
+ \frac{{r_i \xi (\xi  - 2)}}{{12(1 - \xi )^3 }}I_3\nonumber\\
+ \frac{{r_i (3 - 2\xi  - 6r_i \xi  + 2r_i^2 \xi
+ 4r_i \xi ^2  - r_i^2 \xi ^2 )}}{{12(1 - \xi )^3 (r_i  - 1)^2 }}I_4
+\frac{{ r_i(- r_i  - \xi  + 2r_i \xi) }}{{6(r_i  - 1)^2 (1 - \xi )^2 }}I_7\nonumber\\
- \frac{{r_i }}{{12(r_i  - 1)^2 (1 - \xi )}}I_8
- \frac{{r_i \xi ^2 m_W^2 }}{{12(1 - \xi )^2 }}I_9
- \frac{{r_i \xi ^2 m_W^2 }}{{6(1 - \xi )^2 }}I_{11},\nonumber\\
\fl N''=\frac{1}{{2(1 - \xi )^2 }}I_1
- \frac{{r_i }}{{4(r_i  - 1)(1 - \xi )}}I_2
- \frac{{\xi ^2 }}{{2(1 - \xi )^2 }}I_3\nonumber\\
+ \frac{{ - r_i  - 2\xi  + 4r_i \xi  - r_i^2 \xi }} {{4(r_i  -
1)(1 - \xi )^2 }}I_4-\frac{{\xi ^2 }}{{2(1 - \xi )^2 }}I_5 +
\frac{1}{{2(r_i  - 1)(1 - \xi )}}I_8,\nonumber\\
\fl W''=-\frac{\xi^2}{2(\xi-1)^2}I_1+\left[\frac{r_i^2}{6(\xi-1)(r_i-1)^2}-\frac{r_i}{4(r_i-1)(\xi-1)}\right]I_2\nonumber\\
+\left[\frac{r_i\xi(\xi-2)}{12(\xi-1)^3}+\frac{\xi(-r_i+2\xi)}{4(\xi-1)^2}\right]I_3\nonumber\\
+\left[\frac{r_i-2r_i^2+2r_i^3\xi-r_i^3\xi^2}{12(r_i-1)^2(\xi-1)^3}+\frac{r_i+2\xi-4r_i\xi+r_i^2\xi}{4(r_i-1)(\xi-1)^2}\right]I_4+\frac{\xi^2}{2(\xi-1)^2}I_5\nonumber\\
+\left[\frac{1}{2(r_i-1)(\xi-1)}-\frac{r_i}{12(\xi-1)(r_i-1)^2}\right]I_8+\frac{1}{12}m_W^2\frac{r_i\xi^2}{(\xi-1)^2}I_9.\nonumber\\
\end{eqnarray}
Using (A.24),(A.30) and (A.36),we finally obtain the contribution from diagram (a):
\begin{eqnarray}
\fl T(a)=eg_2^2\bar u_\mu(i{\sigma _{\mu \nu }}{q^\nu })\frac{i}{{{{(4\pi
)}^2}}}\frac{1}{{m_W^2}}\Big\{ {m_\mu }A\Big[M(a) + P(a) \ln [r_i] +
Q(a)\ln [\xi ]\Big]\nonumber\\
+ {m_i}B\Big[M'(a) + P'(a)\ln [r_i] + Q'(a)\ln [\xi ]\Big]\nonumber\\
+m_\mu B\Big[M''(a) + P''(a)\ln [r_i]\Big]\nonumber\\
+m_\mu D\Big[M_D(a) + P_D(a)\ln [r_i] + Q_D(a)\ln [\xi ]\Big]\Big\}u_\mu,
\end{eqnarray}
where $M(a),P(a),Q(a),\dots$ are listed in Appendix H.

\clearpage
\section{The calculation of the contribution from Z diagram}
We start with diagram $(b)$:
We begin to calculate the amplitude of diagram Z whose Feynman diagrams are shown in Fig 2.
\begin{eqnarray}
(b)_Z
&=&-\frac{eg_2^2}{4c_W}\ubar_\mu\int_k\frac{(c_{\mu\alpha}^LP_R+c_{\mu\alpha}^RP_L)N_\mu(c_{\alpha\mu}^LP_L+c_{\alpha\mu}^RP_R)}{(k-p)^2-m_Z^2}u_\mu,\\
N_\mu&=&\gamma_\sigma\frac{1}{\kslash-\qslash-m_\alpha}\gamma_\mu\frac{1}{\kslash-m_\alpha}\gamma_\rho\left(g^{\rho\sigma}-(1-\xi)\frac{(k-p)^\rho(k-p)^\sigma}{(k-p)^2-\xi m_Z^2}\right)\nonumber\\
&=&{\rm 1st}+{\rm 2nd},\nonumber\\
c^L_{\alpha\beta}&=&(V^\dagger_LV_L)_{\alpha\beta}-2s_W^2\delta_{\alpha\beta},c^R_{\alpha\beta}=(V^\dagger_RV_R)_{\alpha\beta}-2s_W^2\delta_{\alpha\beta}
\end{eqnarray}
where
\begin{eqnarray}
{\rm 1st}&=&\gamma^\rho\frac{1}{\kslash-\qslash-m_\alpha}\gamma_\mu\frac{1}{\kslash-m_\alpha}\gamma_\rho,\nonumber\\
{\rm 2nd}&=&-(1-\xi)(\kslash-\pslash)\frac{1}{\kslash-\qslash-m_\alpha}\gamma_\mu\frac{1}{\kslash-m_\alpha}(\kslash-\pslash)\frac{1}{(k-p)^2-\xi m_Z^2}.
\end{eqnarray}
Simplify the term 1st.Note that the terms $\gamma_\mu,q_\mu$ can be dropped because they have no contribution to the loop integral.Thus we can get:
\begin{eqnarray}
{\rm 1st}
&\to&\frac{-4k_\mu\kslash+2\kslash\gamma_\mu\qslash+8m_\alpha k_\mu}{\big[(k-q)^2-m_\alpha^2)][k^2-m_\alpha^2\big]}.
\end{eqnarray}
Substituting (B.4) into (B.1) we can get the contribution of the term 1st to the loop integral:
\begin{eqnarray}
T({\rm 1st})
&=&\frac{eg_2^2}{4c_W}\ubar_\mu(i{\sigma _{\mu \nu }}{q^\nu })\big[m_\mu A'(FF)+m_\alpha B'(GG)+m_\mu D'(WW)\big]u_\mu.
\end{eqnarray}
where
\begin{eqnarray}
(FF)&=&\frac{r_\alpha^2}{3(r_\alpha-1)^2}T_{7}+\frac{2r_\alpha}{3(r_\alpha-1)}T_{4}+\frac{1-4r_\alpha}{3(r_\alpha-1)^2}T_{2}+\frac{2}{3(r_\alpha-1)^2}T_{8},\nonumber\\
(GG)&=&\frac{2r_\alpha}{r_\alpha-1}T_{4}-\frac{2}{r_\alpha-1}T_{2},\nonumber\\
(WW)&=&-\frac{r_\alpha^2}{3(r_\alpha-1)^2}T_{7}+\frac{r_\alpha(5-3r_\alpha)}{3(r_\alpha-1)^2}T_{4}+\frac{3r_\alpha-4}{3(r_\alpha-1)^2}T_{2}.
\end{eqnarray}
where $r_\alpha=m_\alpha^2/m_W^2$, and $T_n$ are integral functions listed in
Appendix. 

Now we proceed to the term 2nd.To pick up the revelant terms,again we will use:
\begin{eqnarray}
k_\mu\kslash(k\cdot p)(k\cdot q)
\to\frac{1}{48}(i{\sigma _{\mu \nu }}{q^\nu })\pslash k^4-\frac{1}{48}(i{\sigma _{\mu \nu }}{q^\nu })m_\mu^* k^4,\nonumber\\
k_\mu\kslash(k\cdot p)(k\cdot p)
\to\frac{1}{24}(i{\sigma _{\mu \nu }}{q^\nu })\pslash k^4.
\end{eqnarray}
We can also use the replacement $\pslash\to m_\mu$ when $\pslash$ is on the far right.Thus we can get the contribution of the term 2nd to the loop integral:
\begin{eqnarray}
T({\rm 2nd})&=&\frac{eg_2^2}{4c_W}\ubar_\mu(i{\sigma _{\mu \nu }}{q^\nu })\big[m_\mu A'(FF)'+m_\alpha B'(GG)'+m_\mu D'(WW)'\big]u_\mu.
\end{eqnarray}
where
\begin{eqnarray}
\fl(FF)'=\Big[-\frac{1}{2}\frac{r_\alpha}{(r_\alpha-1)^2}-\frac{5}{3}\frac{r_\alpha}{r_\alpha-1}\Big]T_2-\frac{7}{3}r_\alpha T_3+\Big[\frac{4}{3}\frac{r_\alpha}{r_\alpha-1}+\frac{7}{3}r_\alpha\Big]T_4\nonumber\\
+\Big[\frac{1}{3}\frac{r_\alpha}{r_\alpha-1}+\frac{1}{6}\frac{r_\alpha}{(r_\alpha-1)^2}\Big]T_7+\frac{1}{3}\frac{r_\alpha}{(r_\alpha-1)^2}T_8-\frac{5}{3}m_Z^2r_\alpha\xi T_9-\frac{1}{3}m_Z^4r_\alpha\xi^2T_{10}\nonumber\\
-\frac{1}{3}m_Z^2r_\alpha\xi T_{11}-\frac{1}{6}m_Z^4r_\alpha\xi^2T_{12},\nonumber\\
\fl(GG)'=\frac{3}{2}(\xi-1)T_1+\frac{1}{2(r_\alpha-1)}T_2+\frac{\xi+3r_\alpha}{2}T_3+\Big[-\frac{1}{2}-\frac{3r_\alpha}{2}-\frac{r_\alpha}{2(s_i-1)}\Big]T_4\nonumber\\
+\frac{1}{2}\xi T_5+\frac{1}{2}m_Z^2r_\alpha\xi T_9,\nonumber\\
\fl(WW)'=-\frac{3}{2}(\xi-1)T_1+\frac{-4r_\alpha+3}{6(r_\alpha-1)^2}T_2-\frac{\xi+3r_\alpha}{2}T_3+\Big[\frac{r_\alpha^2}{3(r_\alpha-1)^2}+\frac{1}{2}+\frac{3r_\alpha}{2}\nonumber\\
+\frac{r_\alpha}{2(r_\alpha-1)}\Big]T_4-\frac{1}{2}\xi T_5+\frac{-2r_\alpha^2+r_\alpha}{6(r_\alpha-1)^2}T_7-\frac{1}{2}m_Z^2r_\alpha\xi T_9+\frac{1}{3}m_Z^2r_\alpha\xi T_{11}\nonumber\\
+\frac{1}{6}m_Z^4r_\alpha\xi^2T_{12}.\nonumber\\
\fl A^\prime=\sum_\alpha(c^{L}_{\alpha\mu}c^L_{\alpha\mu}P_R
  +c^{R}_{\alpha\mu}c^R_{\alpha\mu}P_L),\nonumber\\
\fl  B^\prime=\sum_i(c^{L}_{\alpha\mu}c^R_{\alpha\mu}P_R
  +c^{R}_{\alpha\mu}c^L_{\alpha\mu}P_L),\nonumber\\
\fl  D^\prime=\sum_\alpha(c^{L}_{\alpha\mu}c^L_{\alpha\mu}P_L
  +c^{R}_{\alpha\mu}c^R_{\alpha\mu}P_R).
\end{eqnarray}
At last we get the contribution of the diagram $(b)_Z$：
\begin{eqnarray}
\fl T\left[(b)_Z\right]=\frac{eg_2^2}{4c_Wm_Z^2}\frac{i}{(4\pi)^2}\ubar_\mu(i{\sigma _{\mu \nu }}{q^\nu })\Big\{m_\mu A'\big[MM(b) + PP(b) \ln
r_\alpha+QQ(b)\ln\xi\big]\nonumber\\
 +m_\alpha B'\big[MM'(b) + PP'(b)\ln r_\alpha + QQ'(b)\ln\xi\big]\nonumber\\
+m_\mu D'\big[MM_D(b) + PP_D(b)\ln r_\alpha + QQ_D(b)\ln\xi\big]\Big\}u_\mu,
\end{eqnarray}
where $MM(b),PP(b),QQ(b),\dots$ are listed in Appendix I.

Now replacing $Z$ by $G^0$,we get:
\begin{eqnarray}
(b)_{G^0}
&=&-\frac{eg_2^2}{4c_Wm_Z^2}\ubar_\mu\int_k\frac{H}{(k-p)^2-\xi m_Z^2}u_\mu.
\end{eqnarray}
where
\begin{eqnarray}
\fl H=\Big[m_\alpha\big(c_{\mu\alpha}^LP_R+c_{\mu\alpha}^RP_L\big)-m_\mu\big(c_{\mu\alpha}^LP_L+c_{\mu\alpha}^RP_R\big)\Big]N\Big[m_\mu\big(c_{\alpha\mu}^LP_R+c_{\alpha\mu}^RP_L\big)\nonumber\\
-m_\alpha\big(c_{\alpha\mu}^LP_L+c_{\alpha\mu}^RP_R\big)\Big],\nonumber\\
\fl N=\frac{1}{\kslash-\qslash-m_\alpha}\gamma_\mu\frac{1}{\kslash-m_\alpha}.
\end{eqnarray}
Simplify $N$ we get:
\begin{eqnarray}
N&\to&\frac{2k_\mu\kslash+m_\alpha\kslash\gamma_\mu-\qslash\gamma_\mu\kslash-m_\alpha\qslash\gamma_\mu+m_\alpha\gamma_\mu\kslash}{\big[(k-q)^2-m_\alpha^2\big]\big[k^2-m_\alpha^2\big]}
\end{eqnarray}
Substituting (B.13) into (B.11) and using the same method in the diagram b,we can get the contribution of the $b_G^0$ to the amplitude:
\begin{eqnarray}
\fl T[b_{G^0}]
=\frac{eg_2^2}{4c_Wm_Z^2}\frac{i}{(4\pi)^2}\ubar_\mu(i{\sigma _{\mu \nu }}{q^\nu })\Big\{m_\mu A'\big[MM(c) + PP(c) \ln
r_\alpha+QQ(c)\ln\xi\big]\nonumber\\
 +m_\alpha B'\big[MM'(c) + PP'(c)\ln r_\alpha + QQ'(c)\ln\xi\big]\nonumber\\
+m_\mu D'\big[MM_D(c) + PP_D(c)\ln r_\alpha + QQ_D(c)\ln\xi\big]\Big\}u_\mu,
\end{eqnarray}
where $MM(c),PP(c),QQ(c),\dots$ are listed in Appendix I.

\clearpage
\section{The loop functions appearing in Fig 1 are defined
and the results are listed}
\begin{eqnarray*}
\fl I_1=\int {\frac{{{d^4}k}}{{{{(2\pi )}^4}}}} \frac{1}{{({k^2} -
{m^2_i})({k^2} - \xi m^2_W)({k^2} - m^2_W)}}=\frac{i}{{{{(4\pi
)}^2}}}\frac{{(\xi  - 1)r_i\ln r_i - (r_i - 1)\xi \ln \xi }}{{m^2_W(1 -
\xi )(r_i - 1)(r_i - \xi )}},\\
\fl I_2=\int {\frac{{{d^4}k}}{{{{(2\pi )}^4}}}} \frac{1}{{({k^2} -
{m^2_i}){{({k^2} - {m^2_W})}^2}}}=- \frac{i}{{{{(4\pi
)}^2}}}\frac{1}{{{m^2_W}(1 - r_i)}}\Big(1 + \frac{{r_i\ln r_i}}{{1 - r_i}}\Big),\nonumber\\
\fl I_3=\int {\frac{{{d^4}k}}{{{{(2\pi )}^4}}}} \frac{1}{{{{({k^2} -
{m^2_i})}^2}({k^2} - \xi {m^2_W})}}=- \frac{i}{{{{(4\pi
)}^2}}}\frac{1}{{{m^2_W}(r_i - \xi )}}\Big(1 - \frac{{\xi \ln \frac{r_i}{\xi
}}}{{r_i - \xi }}\Big),\nonumber\\
\fl I_4=\int {\frac{{{d^4}k}}{{{{(2\pi )}^4}}}} \frac{1}{{{{({k^2} -
{m^2_i})}^2}({k^2} - {m^2_W})}}=- \frac{i}{{{{(4\pi
)}^2}}}\frac{1}{{{m^2_W}(r_i - 1)}}\Big(1 - \frac{{\ln r_i}}{{r_i - 1}}\Big),\nonumber\\
\fl I_5=\int {\frac{{{d^4}k}}{{{{(2\pi )}^4}}}} \frac{1}{{({k^2} -
{m^2_i}){{({k^2} - \xi {m^2_W})}^2}}}=\frac{i}{{{{(4\pi
)}^2}}}\frac{1}{{{m^2_W}(r_i - \xi )}}\Big(1 + \frac{{r_i\ln \frac{\xi
}{r_i}}}{{r_i - \xi }}\Big),\nonumber\\
\fl I_6=\int {\frac{{{d^4}k}}{{{{(2\pi )}^4}}}} \frac{1}{{{{({k^2} -
\xi {m^2_W})}^3}}}=- \frac{i}{{{{(4\pi
)}^2}}}\frac{1}{2}\frac{1}{{\xi {m^2_W}}},\nonumber\\
\fl I_7=\int {\frac{{{d^4}k}}{{{{(2\pi )}^4}}}} \frac{1}{{{{({k^2} -
{m^2_i})}^3}}}=- \frac{i}{{{{(4\pi
)}^2}}}\frac{1}{2}\frac{1}{{r_i{m^2_W}}},\nonumber\\
\fl I_8=\int {\frac{{{d^4}k}}{{{{(2\pi )}^4}}}} \frac{1}{{{{({k^2} -
{m^2_W})}^3}}}=- \frac{i}{{{{(4\pi
)}^2}}}\frac{1}{2}\frac{1}{{{m^2_W}}},\nonumber\\
\fl I_9=\int {\frac{{{d^4}k}}{{{{(2\pi)}^4}}}} \frac{1}{{{{({k^2} -
{m^2_i})}^2}{({k^2}-\xi m_W^2)^2}}}
=-\frac{i}{(4\pi)^2}\frac{1}{m_W^4(r_i-\xi)^2}\Big(2 +\frac{{(r_i + \xi
)\ln \frac{\xi }{r_i}}}{{{{(r_i - \xi )}}}}\Big),\\
\fl I_{10}=\int {\frac{{{d^4}k}}{{{{(2\pi )}^4}}}} \frac{1}{{{{({k^2}
- {m^2_i})}^2}{{({k^2} - \xi {m^2_W})}^3}}} %
=\frac{i}{{{{(4\pi)}^2}{m^6_W}{{(r_i - \xi )}^3}}}\Big( - \frac{{r_i +
5\xi }}{{2\xi }} - \frac{{2r_i + \xi }}{{r_i - \xi }}\ln \frac{\xi }{r_i}\Big),\nonumber\\
\fl I_{11}=\int {\frac{{{d^4}k}}{{{{(2\pi )}^4}}}} \frac{1}{{{{({k^2}
- {m^2_i})}^3}({k^2} - \xi {m^2_W})}}%
=\frac{i}{{{{(4\pi )}^2}{m_W}^4{{(r_i - \xi )}^2}}}\Big(\frac{1}{2} +
\frac{\xi }{{2r_i}} - \frac{{\xi \ln \frac{r_i}{\xi }}}{{r_i - \xi }}\Big),\nonumber\\
\fl I_{12}=\int {\frac{{{d^4}k}}{{{{(2\pi )}^4}}}} \frac{1}{{{{({k^2}
- {m^2_i})}^3}{{({k^2} - \xi {m^2_W})}^2}}}%
=\frac{i}{{{{(4\pi)}^2}{m^6_W}{{(r_i - \xi )}^3}}}\Big(\frac{{5r_i + \xi
}}{{2r_i}} + \frac{{r_i + 2\xi }}{{r_i - \xi }}\ln \frac{r_i}{\xi }\Big),\nonumber\\
\fl I_{13}=\int {\frac{{{d^4}k}}{{{{(2\pi )}^4}}}} \frac{1}{{({k^2} -
{m_i}^2){{({k^2} - \xi {m^2_W})}^3}}}%
=\frac{i}{{{{(4\pi )}^2}}}\frac{1}{{2{m^4_W}{{(r_i - \xi )}^2}}}\Big(1 +
\frac{r_i}{\xi } + \frac{{2r_i\ln \frac{\xi }{r_i}}}{{r_i - \xi }}\Big).
\end{eqnarray*}

\clearpage
\section{The loop functions appearing in Fig 2 are defined
and the results are listed.}
\begin{eqnarray*}
\fl T_1=\int {\frac{{{d^4}k}}{{{{(2\pi )}^4}}}} \frac{1}{{({k^2} -
{m^2_\alpha})({k^2} - \xi {m^2_Z})({k^2} - {m^2_Z})}}=\frac{i}{{{{(4\pi
)}^2}}}\frac{{(\xi  - 1)r_\alpha\ln r_\alpha - (r_\alpha - 1)\xi \ln \xi }}{{{m^2_Z}(1 -
\xi )(r_\alpha - 1)(r_\alpha - \xi )}},\\
\fl T_2=\int {\frac{{{d^4}k}}{{{{(2\pi )}^4}}}} \frac{1}{{({k^2} -
{m^2_\alpha}){{({k^2} - {m^2_Z})}^2}}}=- \frac{i}{{{{(4\pi
)}^2}}}\frac{1}{{{m^2_Z}(1 - r_\alpha)}}\Big(1 + \frac{{r_\alpha\ln r_\alpha}}{{1 - r_\alpha}}\Big),\nonumber\\
\fl T_3=\int {\frac{{{d^4}k}}{{{{(2\pi )}^4}}}} \frac{1}{{{{({k^2} -
{m^2_\alpha})}^2}({k^2} - \xi {m^2_Z})}}=- \frac{i}{{{{(4\pi
)}^2}}}\frac{1}{{{m^2_Z}(r_\alpha - \xi )}}\Big(1 - \frac{{\xi \ln \frac{r_\alpha}{\xi
}}}{{r_\alpha - \xi }}\Big),\nonumber\\
\fl T_4=\int {\frac{{{d^4}k}}{{{{(2\pi )}^4}}}} \frac{1}{{{{({k^2} -
{m^2_\alpha})}^2}({k^2} - {m^2_Z})}}=- \frac{i}{{{{(4\pi
)}^2}}}\frac{1}{{{m^2_Z}(r_\alpha - 1)}}\Big(1 - \frac{{\ln r_\alpha}}{{r_\alpha - 1}}\Big),\nonumber\\
\fl T_5=\int {\frac{{{d^4}k}}{{{{(2\pi )}^4}}}} \frac{1}{{({k^2} -
{m^2_\alpha}){{({k^2} - \xi {m^2_Z})}^2}}}=\frac{i}{{{{(4\pi
)}^2}}}\frac{1}{{{m^2_Z}^2(r_\alpha - \xi )}}\Big(1 + \frac{{r_\alpha\ln \frac{\xi
}{r_\alpha}}}{{r_\alpha - \xi }}\Big),\nonumber\\
\fl T_6=\int {\frac{{{d^4}k}}{{{{(2\pi )}^4}}}} \frac{1}{{{{({k^2} -
\xi {m^2_Z})}^3}}}=- \frac{i}{{{{(4\pi
)}^2}}}\frac{1}{2}\frac{1}{{\xi {m^2_Z}}},\nonumber\\
\fl T_7=\int {\frac{{{d^4}k}}{{{{(2\pi )}^4}}}} \frac{1}{{{{({k^2} -
{m^2_\alpha})}^3}}}=- \frac{i}{{{{(4\pi
)}^2}}}\frac{1}{2}\frac{1}{{r_\alpha{m^2_Z}}},\nonumber\\
\fl T_8=\int {\frac{{{d^4}k}}{{{{(2\pi )}^4}}}} \frac{1}{{{{({k^2} -
{m^2_Z})}^3}}}=- \frac{i}{{{{(4\pi
)}^2}}}\frac{1}{2}\frac{1}{{{m^2_Z}}},\nonumber\\
\fl T_9=\int {\frac{{{d^4}k}}{{{{(2\pi)}^4}}}} \frac{1}{{{{({k^2} -
{m^2_\alpha})}^2}{({k^2}-\xi m_Z^2)^2}}}
=-\frac{i}{(4\pi)^2}\frac{1}{m_Z^4(r_\alpha-\xi)^2}\Big(2 +\frac{{(r_\alpha + \xi
)\ln \frac{\xi }{r_\alpha}}}{{{{(r_\alpha - \xi )}}}}\Big),\\
\fl T_{10}=\int {\frac{{{d^4}k}}{{{{(2\pi )}^4}}}} \frac{1}{{{{({k^2}
- {m^2_\alpha})}^2}{{({k^2} - \xi {m^2_Z})}^3}}} %
=\frac{i}{{{{(4\pi)}^2}{m^6_Z}{{(r_\alpha - \xi )}^3}}}\Big( - \frac{{r_\alpha +
5\xi }}{{2\xi }} - \frac{{2r_\alpha + \xi }}{{r_\alpha - \xi }}\ln \frac{\xi }{r_\alpha}\Big),\nonumber\\
\fl T_{11}=\int {\frac{{{d^4}k}}{{{{(2\pi )}^4}}}} \frac{1}{{{{({k^2}
- {m^2_\alpha})}^3}({k^2} - \xi {m^2_Z})}}%
=\frac{i}{{{{(4\pi )}^2}{m^4_Z}{{(r_\alpha - \xi )}^2}}}\Big(\frac{1}{2} +
\frac{\xi }{{2r_\alpha}} - \frac{{\xi \ln \frac{r_\alpha}{\xi }}}{{r_\alpha - \xi }}\Big),\nonumber\\
\fl T_{12}=\int {\frac{{{d^4}k}}{{{{(2\pi )}^4}}}} \frac{1}{{{{({k^2}
- {m^2_\alpha})}^3}{{({k^2} - \xi {m^2_Z})}^2}}}%
=\frac{i}{{{{(4\pi)}^2}{m^6_Z}{{(r_\alpha - \xi )}^3}}}\Big(\frac{{5r_\alpha + \xi
}}{{2r_\alpha}} + \frac{{r_\alpha + 2\xi }}{{r_\alpha - \xi }}\ln \frac{r_\alpha}{\xi }\Big),\nonumber\\
\fl T_{13}=\int {\frac{{{d^4}k}}{{{{(2\pi )}^4}}}} \frac{1}{{({k^2} -
{m^2_\alpha}){{({k^2} - \xi {m^2_Z})}^3}}}%
=\frac{i}{{{{(4\pi )}^2}}}\frac{1}{{2{m^4_Z}{{(r_\alpha - \xi )}^2}}}\Big(1 +
\frac{r_\alpha}{\xi } + \frac{{2r_\alpha\ln \frac{\xi }{r_\alpha}}}{{r_\alpha - \xi }}\Big).
\end{eqnarray*}

\clearpage
\section{The loop functions appearing in diagram(b) are listed.}
\begin{eqnarray*}
\fl M(b)=\frac{{- {\rm{5  +  r_i  +  5}}\xi  - {\rm{2
r_i}}\xi }}{{{{24(\xi  - 1)}^2}(r_i - 1)}} + \frac{{r_i - {r_i^2} + \xi  +
2{r_i^2}\xi  - {\xi ^2} - 2r_i{\xi ^2}}}{{{{24(\xi  - 1)}^2}{{(r_i - \xi
)}^2}}}\nonumber\\
+\frac{{ - 2{r_i^2} - 7{r_i^2}\xi  + 2{\xi ^2} + 13r_i{\xi
^2}}}{{{{48(r_i - \xi )}^3}}},\nonumber\\
\fl P(b)=r_i\Big[\frac{{\xi }}{{4(\xi  - 1)(r_i - 1)(r_i - \xi )}}
- \frac{{-4+3\xi}}{{{{24(\xi  - 1)}^2}{{(r_i - 1)}^2}}}\nonumber\\
+\frac{\xi(-2+r_i+\xi-2r_i\xi+2\xi^2)}{24(r_i-\xi)^3(\xi-1)^2}-\frac{{{\xi ^3}}}{{{{8(r_i - \xi )}^4}}}\Big],\nonumber\\
\fl Q(b)=\frac{\xi}{(r_i-\xi)^2}\left[\frac{{ - r_i - 4r_i\xi  + 6{\xi
^2}}}{24(\xi  - 1)^2} + \frac{{r_i (r_i - \xi -
r_i\xi )}}{{12(\xi  - 1){{(r_i - \xi )}}}} + \frac{{r_i{\xi ^2}}}{{{{8(r_i -
\xi )}^2}}}\right],\nonumber\\
\fl M'(b)=\frac{1}{r_i-\xi}\left[\frac{{ - r_i + 2\xi  - r_i\xi }}{{8(\xi
- 1)(r_i - 1)}} + \frac{{r_i\xi }}{4(r_i - \xi )}\right],\nonumber\\
\fl P'(b)=\frac{r_i}{{{{(r_i - \xi )}^2}}}\left[\frac{{{r_i^2} + \xi  - 2r_i\xi
- {r_i^2}\xi  - {\xi ^2} + 2r_i{\xi ^2}}}{{{{8(r_i - 1)}^2}(\xi  - 1)}} -
\frac{{\xi (r_i + \xi )}}{8(r_i - \xi )}\right],\nonumber\\
\fl Q'(b)=\frac{\xi}{{{{(r_i - \xi )}^2}}}\left[\frac{{r_i + r_i\xi
- 2{\xi ^2}}}{{{{8(\xi  - 1)}^2}}} + \frac{{r_i (r_i + \xi
)}}{8(r_i - \xi )}\right],\nonumber\\
\fl M_D(b)=\frac{r_i}{r_i-\xi}\left[\frac{1}{8(r_i-1)}-\frac{\xi}{4(r_i-\xi)}\right]+\frac{r_i}{12(\xi-1)(r_i-1)^2}+\frac{r_i(1-2\xi)}{24(r_i-1)(r_i-\xi)(\xi-1)}\nonumber\\
+\frac{r_i\xi(3r_i-\xi)}{16(r_i-\xi)^3},\nonumber\\
\fl P_D(b)=\frac{r_i}{(r_i-\xi)^2}\left[\frac{-r_i^2+\xi}{8(r_i-1)^2}+\frac{\xi(r_i+\xi)}{8(r_i-\xi)}\right]+\frac{r_i(2r_i-3r_i\xi+\xi)}{24(r_i-1)^3(\xi-1)^2}+\frac{r_i\xi(2\xi-1)}{24(\xi-1)^2(r_i-\xi)^2}\nonumber\\
+\frac{r_i\xi(-2r_i^2-2r_i\xi+\xi^2)}{24(r_i-\xi)^4},\nonumber\\
\fl Q_D(b)=\frac{r_i\xi}{(r_i-\xi)^2}\left[\frac{1}{8(\xi-1)}-\frac{r_i+\xi}{8(r_i-\xi)}\right]+\frac{r_i\xi}{(r_i-\xi)^2}\left[\frac{-2\xi+1}{24(\xi-1)^2}+\frac{2r_i^2+2r_i\xi-\xi^2}{24(r_i-\xi)^2}\right].\nonumber\\
\end{eqnarray*}

\clearpage
\section{The loop functions appearing in diagram(c) are listed.}
\begin{eqnarray*}
\fl M(c)=\frac{{ - {r_i^2} - r_i\xi  + 2{\xi ^2} + 3r_i{\xi
^2} - 3{\xi ^3} - r_i{\xi ^3} + {\xi ^4}}}{{{{24(r_i - \xi )}^2}{{(\xi  -
1)}^2}}} + \frac{{ - 1 - 2r_i + {r_i^2} + \xi  + r_i\xi }}{{{{24(r_i -
1)}^2}{{(\xi  - 1)}^2}}},\nonumber\\
+\frac{{\xi ( - {r_i^2} + 5r_i\xi  + 2{\xi ^2})}}{{{{48(r_i - \xi )}^3}}},
\nonumber\\
\fl P(c)=r_i\Big[\frac{{3 - r_i - 2\xi }}{{24(\xi  - 1)^2 (r_i - 1)^3 }}
+ \frac{1}{{24(r_i - \xi )^2 (\xi  - 1)^2 }}\nonumber\\
 + \frac{{ - r_i^2  + 2r_i\xi  + r_i^2 \xi  - \xi ^2  - 2r_i\xi ^2  - 2\xi ^3 }}
{{24(r_i - \xi )^4 }}\Big],\nonumber\\
\fl Q(c)=-P(c) + \frac{{r_i(3 - r_i - 2\xi )}}{{24{{(\xi  - 1)}^2}{{(r_i -
1)}^3}}},\nonumber\\
\fl M'(c)=\frac{1}{{r_i - \xi }}\Big[\frac{{ - r_i\xi  - r_i + 2\xi }}{{8(r_i -
1)(\xi  - 1)}} + \frac{{r_i\xi }}{4({r_i - \xi })}\Big],\nonumber\\
\fl P'(c)=\frac{{r_i(1 - 2\xi )}}{{4(r_i - \xi )(r_i - 1)(\xi -
1)}} + \frac{{\xi ( - 2\xi  + 3r_i)}}{{8(\xi  - 1){{(r_i - \xi )}^2}}}\nonumber\\
+\frac{{2 - r_i}}{{8(\xi  - 1){{(r_i - 1)}^2}}} - \frac{{r_i\xi (r_i + \xi
)}}{{{{8(r_i - \xi )}^3}}},\nonumber\\
\fl Q'(c)=\frac{\xi}{{{{(r_i - \xi )}^2}}}\Big[\frac{{r_i + r_i\xi
- 2{\xi ^2}}}{{{{8(\xi  - 1)}^2}}} + \frac{{r_i(r_i + \xi
)}}{8({r_i - \xi })}\Big],\nonumber\\
\fl M_D(c)=\frac{1}{r_i-\xi}\Big[\frac{4r_i-6\xi+3r_i\xi}{24(r_i-1)(\xi-1)}-\frac{r_i\xi}{4(r_i-\xi)}\Big]+\frac{r_i\xi(3\xi+\xi^2-9r_i+5r_i\xi)}{48(r_i-\xi)^3(\xi-1)},\nonumber\\
\fl P_D(c)=\frac{1}{\xi-1}\Big[-\frac{(-2\xi+1)r_i}{4(r_i-1)(r_i-\xi)}+\frac{r_i-2}{8(r_i-1)^2}+\frac{\xi(-4r_i^2+r_i^2\xi-2\xi^2+4r_i\xi+r_i\xi^2)}{8(r_i-\xi)^3}\Big]\nonumber\\
+\frac{r_i}{(\xi-1)^2}\Big[\frac{1}{24(r_i-1)^2}-\frac{\xi(-4r_i\xi^2+2r_i\xi^3+\xi^3+3r_i^2-3r_i^2\xi+r_i^2\xi^2)}{24(r_i-\xi)^4}\Big],\nonumber\\
\fl Q_D(c)=\frac{\xi}{(r_i-\xi)^2}\Big[\frac{-r_i-r_i\xi+2\xi^2}{8(\xi-1)^2}-\frac{r_i(r_i+\xi)}{8(r_i-\xi)}\Big]\nonumber\\
+\frac{r_i\xi(-4r_i\xi^2+2r_i\xi^3+\xi^3+3r_i^2-3r_i^2\xi+r_i^2\xi^2)}{24(\xi-1)^2(r_i-\xi)^4}.
\end{eqnarray*}

\clearpage
\section{The loop functions appearing in diagram(d) are listed.}
\begin{eqnarray*}
M(d)&=&\frac{1}{{24}}\frac{{r_i(4{r_i^2} - 5r_i\xi  - 5{\xi ^2})}}{{{{(r_i -
\xi )}^3}}},\nonumber\\
P(d)&=&-Q(d)=\frac{{{r_i^2}\xi ( - r_i + 2\xi )}}{{4{{(r_i - \xi )}^4}}},
\nonumber\\
M'(d)&=&\frac{{ - r_i(\xi+r_i)}}{{{{4(r_i - \xi )}^2}}},
\nonumber\\
P'(d)&=&-Q'(d)=\frac{{{r_i^2}\xi }}{{2{{(r_i - \xi )}^3}}},\nonumber\\
M_D(d)&=&\frac{r_i(r_i+\xi)}{4(r_i-\xi)^2}+\frac{r_i(-2r_i^2-5r_i\xi+\xi^2)}{24(r_i-\xi)^3},\nonumber\\
P_D(d)&=&-Q_D(d)=\frac{r_i^3\xi}{4(r_i-\xi)^4}-\frac{r_i^2\xi}{2(r_i-\xi)^3}.
\end{eqnarray*}

\clearpage
\section{The loop functions appearing in diagram(a) are listed.}
\begin{eqnarray*}
\fl M(a)=\frac{{8 - 35r_i + 66{r_i^2} - 41{r_i^3} +
2{r_i^4}}}{{{{24(r_i - 1)}^4}}} - \frac{{3r_i - 2}}{{{{12(r_i - 1)}^2}(\xi  -
1)}}\nonumber\\
 - \frac{{r_i}}{{12(r_i - \xi )(r_i - 1)}}+\frac{{{r_i^2} - r_i\xi  -
{r_i^2}\xi  + r_i{\xi ^2} + {\xi ^3}}}{{{{12(r_i - \xi )}^2}(\xi  - 1)}},\nonumber\\
\fl P(a)=r_i\Big[\frac{{ - \xi }}{{4(\xi  - 1)(r_i - \xi )(r_i -
1)}} + \frac{{3{r_i^2}}}{{{{4(r_i - 1)}^4}}}\nonumber\\
 + \frac{{3r_i - 1}}{{{{24(r_i -
1)}^3}(\xi  - 1)}} + \frac{r_i-3\xi+2r_i\xi-2\xi^2}{{24(\xi  -
1){{(r_i - \xi )}^3}}}+\frac{1+5\xi}{24(r_i-\xi)^2}\Big],\nonumber\\
\fl Q(a)=\frac{{{\xi ^2}}}{{(\xi  - 1){{(r_i - \xi )}^2}}}\Big[\frac{{4r_i
- 3\xi  - r_i\xi }}{12({\xi  - 1})} + \frac{{r_i( - r_i + 3\xi )}}{{24(r_i - \xi
)}}\Big],\nonumber\\
\fl M'(a)=\frac{1}{{r_i - 1}}\Big[\frac{{r_i{\xi ^2} - {\xi ^2} + r_i - \xi
}}{{4(\xi  - 1)(r_i - \xi )}} + \frac{{3(3r_i - 1)}}{4({r_i - 1})}\Big],\nonumber\\
\fl P'(a)=\frac{r_i^2}{(1-r_i)^2}\Big[-\frac{{r_i(\xi  - 1)}}{{4(r_i - \xi )^2 }}+\frac{3}{2(1-r_i)}\Big],\nonumber\\
\fl Q'(a)=\frac{1}{4}\frac{{{\xi ^2}( - 3r_i + 2\xi  + r_i\xi
)}}{{{{(\xi - 1)}^2}{{(r_i - \xi )}^2}}},\nonumber\\
\fl M''(a)=\frac{11-13r_i}{4(r_i-1)^2},\nonumber\\
\fl P''(a)=\frac{r_i(-5+6r_i)}{2(r_i-1)^3},\nonumber\\
\fl M_D(a)=\frac{3r_i-1}{8(r_i-1)^2}+
\frac{2r_i^2+5r_i-1}{6(r_i-1)^3}-\frac{\xi^2}{4(\xi-1)(r_i-\xi)}\nonumber\\
+\frac{r_i}{\xi-1}\left[\frac{-1+7r_i-5r_i\xi+\xi-2r_i^2\xi)}{24(r_i-1)^3}+\frac{\xi}{12(r_i-\xi)}+\frac{\xi^2}{12(r_i-\xi)^2}\right],\nonumber\\
\fl P_D(a)=-\frac{r_i^2}{4(r_i-1)^3}-\frac{r_i^2}{(r_i-1)^4}+\frac{\xi(-r_i+\xi-r_i\xi)}{4(\xi-1)(r_i-1)(r_i-\xi)}+\frac{(2r_i-\xi)\xi^2}{4(r_i-\xi)^2(\xi-1)}\nonumber\\
+\frac{r_i}{\xi-1}\left[\frac{-4r_i-3r_i^2+1+6r_i^2\xi}{24(r_i-1)^4}-\frac{\xi^2}{12(r_i-\xi)^2}-\frac{\xi^2(r_i+\xi)}{24(r_i-\xi)^3}\right],\nonumber\\
\fl Q_D(a)=\frac{r_i\xi^2}{12(\xi-1)(r_i-\xi)^2}+\frac{r_i\xi^2(r_i+\xi)}{24(\xi-1)(r_i-\xi)^3}-\frac{\xi^2(-2r_i+r_i\xi+\xi)}{4(r_i-\xi)^2(\xi-1)^2}.
\end{eqnarray*}

\clearpage
\section{The loop functions appearing in diagram $Z$ are listed.}
\begin{eqnarray*}
\fl MM(b)=\frac{7-3r_\alpha+6r_\alpha^2-28r_\alpha^3}{12(r_\alpha-1)^3}+\frac{30r_\alpha^3-22r_\alpha^2\xi-r_\alpha\xi^2-\xi^3}{12(r_\alpha-\xi)^3},\nonumber\\
\fl PP(b)=\frac{r_\alpha^2(-5+8r_\alpha)}{2(r_\alpha-1)^4}-\frac{r_\alpha\xi(24r_\alpha^2-25r_\alpha\xi+10\xi^2)}{6(r_\alpha-\xi)^4},\nonumber\\
\fl QQ(b)=\frac{r_\alpha\xi(24r_\alpha^2-25r_\alpha\xi+10\xi^2)}{6(r_\alpha-\xi)^4},\nonumber\\
\fl MM'(b)=\frac{-4-5r_\alpha+3r_\alpha^2}{2(r_\alpha-1)^2}+\frac{r_\alpha(-3r_\alpha+\xi)}{2(r_\alpha-\xi)^2},\nonumber\\
\fl PP'(b)=\frac{1-3r_\alpha-3r_\alpha^2-r_\alpha^3}{2(r_\alpha-1)^3}+\frac{r_\alpha^3+r_\alpha^2\xi+r_\alpha\xi^2-\xi^3}{2(r_\alpha-\xi)^3},\nonumber\\
\fl QQ'(b)=-\frac{r_\alpha\xi(-2r_\alpha+\xi)}{(r_\alpha-\xi)^3},\nonumber\\
\fl MM_D(b)=-\frac{5+7r_\alpha-12r_\alpha^2+6r_\alpha^3}{4(r_\alpha-1)^3}+\frac{18r_\alpha^3-22r_\alpha^2\xi+11r_\alpha\xi^2-\xi^3}{12(r_\alpha-\xi)^3},\nonumber\\
\fl PP_D(b)=\frac{-1+4r_\alpha-9r_\alpha^2+4r_\alpha^3-r_\alpha^4}{6(r_\alpha-1)^4}+\frac{3r_\alpha^4-12r_\alpha^3\xi+21r_\alpha^2\xi^2-10r_\alpha\xi^3+\xi^4}{6(r_\alpha-\xi)^4},\nonumber\\
\fl QQ_D(b)=\frac{r_\alpha(-2r_\alpha^3+8r_\alpha^2\xi-15r_\alpha\xi^2+6\xi^3)}{6(r_\alpha-\xi)^4},\nonumber\\
\fl MM(c)=\frac{r_\alpha(-29r_\alpha^2+19r_\alpha\xi+4\xi^2)}{12(r_\alpha-\xi)^3},\nonumber\\
\fl PP(c)=-QQ(c)=\frac{r_\alpha\xi(24r_\alpha^2-25r_\alpha\xi+10\xi^2)}{6(r_\alpha-\xi)^4},\nonumber\\
\fl MM'(c)=\frac{r_\alpha(3r_\alpha-\xi)}{2(r_\alpha-\xi)^2},\nonumber\\
\fl PP'(c)=-QQ'(c)=\frac{r_\alpha\xi(-2r_\alpha+\xi)}{(r_\alpha-\xi)^3},\nonumber\\
\fl MM_D(c)=-\frac{r_\alpha(17r_\alpha^2-19r_\alpha\xi+8\xi^2)}{12(r_\alpha-\xi)^3},\nonumber\\
\fl PP_D(c)=-QQ_D(c)=\frac{r_\alpha(-2r_\alpha^3+8r_\alpha^2\xi-15r_\alpha\xi^2+6\xi^3)}{6(r_\alpha-\xi)^4}.
\end{eqnarray*}

\end{document}